\documentstyle[12pt,aaspp4]{article}
\newcommand{\etal}{{\it{et al.}}~}

\newcommand{\eg}{{\it{e.g.}}}	
\newcommand{\da}{{\tablenotemark{\dagger}}}
\begin{document}
\title{Spectrophotometry of HII Regions, Diffuse Ionized Gas and 
Supernova Remnants in M31: The Transition from Photo- to
Shock-Ionization}

\author{Vanessa C. Galarza and Ren\'e A.M. Walterbos\altaffilmark{1}}
\affil{New Mexico State University, Department of Astronomy, Box 4500, 
Dept 30001, Las Cruces, NM 88003}
\and
\author{Robert Braun\altaffilmark{1}}
\affil{Netherlands Foundation for Research in Astronomy, 
P.O. Box 2, 7990AA Dwingeloo, The Netherlands}

\altaffiltext{1}{Visiting Astronomer, Kitt Peak National Observatory,
National Optical Astronomy Observatories, which is operated by the
Association of Universities for Research in Astronomy, Inc. (AURA)
under cooperative agreement with the National Science Foundation. }

\begin{abstract}

We present results of KPNO 4-m optical spectroscopy of discrete
emission-line nebulae and regions of diffuse ionized gas (DIG) in M31.
Long-slit spectra of 16 positions in the NE half of M31 were obtained
over a 5-15 kpc range in radial distance from the center of the galaxy
with special attention to the annulus of high star formation between 8
and 12 kpc.  Slit positions were chosen such that several
emission-line nebulae and large sections of diffuse gas could be
studied, without compromising the parallactic angle.  The spectra have
been used to confirm 16 supernova remnant candidates from the
\markcite{bw93}Braun \& Walterbos (1993) catalog.  The slits also
covered 46 HII regions which show significant differences among the
various morphological types (center-brightened, diffuse, rings).
Radial gradients in emission-line ratios such as [OIII]/H$\beta$ and
[OII]/[OIII] are observed most prominently in the center-brightened
HII regions.  These line ratio trends are either much weaker or
completely absent in the diffuse and ring nebulae. The line ratio
gradients previously seen in M31 SNRs (\markcite{bkc81}Blair,
Kirshner, \& Chevalier 1981;
\markcite{bkc82}Blair, Kirshner, \& Chevalier 1982) are well
reproduced by our new data. The spectra of center-brightened HII
regions and SNRs confirm previous determinations of the radial
abundance gradient in M31.  We use diagnostic diagrams which separate
photoionized gas from shock-ionized gas to compare the spectral
properties of HII regions, SNRs and DIG. This analysis strengthens
earlier claims (\markcite{gwb97}Greenawalt, Walterbos, \& Braun 1997)
that the DIG in the disk of M31 is photoionized by a dilute radiation
field.

\end{abstract}


\section{Introduction}

Our nearest large galaxy neighbor, M31, provides us with an excellent
opportunity to study the star-formation process and the properties of
the interstellar medium in great detail.  This Sb spiral is
interesting in that it has a low star formation rate, about
0.2$\sim$0.5 M$_{\odot}$/yr (\markcite{w88}Walterbos 1988), relative
to the typical Sb galaxies, which can have star formation rates up to
4M$_{\odot}$/yr (\markcite{ktc94}Kennicutt, Tamblyn, \& Congdon 1994).
M31's proximity\footnotemark[1] has inspired numerous studies of its
emission-line nebulae both in imaging and spectroscopic mode
(e.g. \markcite{pavcmms78}Pellet \etal 1978; \markcite{bkc81}Blair,
Kirshner, \& Chevalier 1981; \markcite{dk81}Dennefeld \& Kunth 1981;
\markcite{bkc82}Blair, Kirshner, \& Chevalier 1982; 
\markcite{dbdb84}Dopita \etal 1984a; 
\markcite{wb92}Walterbos \& Braun 1992; 
\markcite{mla93}Meyssonnier, Lequeux, \& Azzopardi 1993; 
\markcite{mpplshpt95}Magnier \etal 1995). Due to the large angular 
size of M31, the spatial coverage of the spectroscopic observations
has necessarily been rather limited.

\footnotetext[1]{\markcite{fm90}Freedman \& Madore 1990 determined a 
distance for M31 $\sim$ 750 kpc based on Cepheid observations; for
this work, however, we adopt the traditional value D $\sim$ 690 kpc,
consistent with the \markcite{wb92}WB92 catalog.}

An H$\alpha$ and [SII] 6716,31$\rm\AA$ imaging survey of most of the 
northeast half of M31 resulted in a catalog of 958 gaseous nebulae
(\markcite{wb92}Walterbos \& Braun 1992, hereafter
\markcite{wb92}WB92).  The sensitive survey (rms noise $\sim$ 1.1 x
10$^{-17}$ erg cm$^{-2}$ sec$^{-1}$ pix$^{-1}$) also allowed
\markcite{wb94}Walterbos \& Braun (1994) to study the diffuse ionized
gas (DIG) in this galaxy.  This faint component of the ISM, which
permeates the disk and can contribute up to 50\% of the total
H$\alpha$ luminosity of spiral galaxies (see also 
\eg \markcite{fwgh96}Ferguson \etal 1996; 
\markcite{cwb96}Hoopes \etal 1996) remains ill-understood
because the ionizing source has not been well constrained.  The
proximity of M31 facilitates the isolation of gaseous nebulae from the
DIG, allowing a direct comparison of their properties.

The \markcite{wb92}WB92 survey also yielded a set of 52 supernova
remnant (SNR) candidates, presented in \markcite{bw93}Braun \&
Walterbos (1993, hereafter \markcite{bw93}BW93), which are generally
fainter than the candidates in other M31 surveys such as
\markcite{bkc81}Blair, Kirshner, \& Chevalier 1981 and
\markcite{mpvlshpt95}Magnier \etal 1995.  To confirm the
\markcite{bw93}BW93 SNR candidates, and to further study and compare
the spectral properties of HII regions and DIG in M31, we chose
several targets for spectroscopic observation.  This paper presents
the optical spectrophotometric results of a sample of 44 distinct HII
regions, 18 SNR candidates (two of which turned out to be HII
regions), and numerous regions of DIG.
\markcite{gwb}Greenawalt, Walterbos, \& Braun (1997) discussed the
global spectral characteristics of the DIG in M31 based on averaged
spectra obtained from the same data discussed here. In this paper, we
will compare spectra of individual regions of DIG with HII region and
SNR spectra.

Details about the observations and data reduction are given in \S 2.
We discuss the extinction corrections in \S 3.  Next we consider the
nebular conditions and radial trends in various line ratios for our
sample of HII regions in \S 4.  In \S 5, we discuss the confirmation
and radial spectral line variations of the SNRs. This section
concludes with a summary of what we can determine about the radial
abundance gradient in M31 from the HII region and SNR spectra. It will
become clear that while abundance trends are definitely present, line
ratio variations due to excitation effects in HII regions and
variations in shock conditions in SNRs contribute substantial scatter
to the line ratios. We therefore conclude with a comparison of the
emission-line properties of HII regions, DIG and SNRs in \S 6, to shed
further light on the ionization mechanism for the DIG.  A summary of
our results is presented in \S 7.

\section{Observations and Data Reduction}

A complete description of the observations and data reduction has
already been provided elsewhere (\markcite{gwb97}Greenawalt,
Walterbos, \& Braun 1997); we briefly summarize the important aspects
here.

Our long-slit spectra were obtained with the RC spectrograph on the
Mayall 4m telescope at Kitt Peak National Observatory in November
1991.  The dispersion of the KPC-10A grating we used is
2.77$\rm\AA$/pixel which provides a spectral resolution of 6$\rm\AA$.
The total spectral coverage was 3550$\rm\AA$ to 6850$\rm\AA$ with a
large overlap region 4000-6400$\rm\AA$.  The dimensions of the slit
were 2'' X 5.4', with a spatial scale of 0.69''/pixel.

Sixteen different slit positions were obtained on the NE half of M31,
concentrated in the three spirals arms between 5 kpc and 15 kpc from
the center of the galaxy, with special attention to the annulus of
high star-formation activity near 10 kpc.  The slit positions on the
sky were carefully chosen to obtain at least one HII region, one SNR
candidate, and DIG in one slit.  Two 15-minute observations were taken
at each position with an observation of a nearby standard star in
between.  These standards aided in the flux-calibration of the object
spectra, which was conducted using the standard IRAF\footnotemark[2]
package techniques. Table 1 provides the following details on the
observations: 2D spectrum name which corresponds to the
\markcite{wb92}WB92 identification for the most prominent source in
the slit (column 1); epoch 2000 right-ascension and declination for
slit center (columns 2 and 3); slit position angle on the sky chosen
in each case such as not to compromise the parallactic angle (column
4); image field from \markcite{wb92}WB92 where the slit was positioned
(column 5); and all \markcite{wb92}WB92 catalog objects for which 1D
spectra were extracted (column 6).

\footnotetext[2]{IRAF is distributed by the National Optical Astronomy 
Observatories, which is operated by the Association for Research in 
Astronomy, Inc., under cooperative agreement with the National 
Science Foundation.}

The extraction of one-dimensional spectra from the two-dimensional
images proceeded as follows.  All discrete objects on the 2D spectra
were identified on the H$\alpha$ images of \markcite{wb92}WB92.  The
object boundaries were determined from a comparison of the H$\alpha$
emission-line profile along the slit and a cut across the H$\alpha$
images which most closely corresponds to the position of the slit on
the sky.  This was not always a straight-forward task because of the
difference in spatial resolution between the spectra and the
emission-line images, but close matches were obtained after some trial
and error.  Sky apertures were identified as close as possible on
either side of the object apertures and often included the faint
emission from nearby DIG.  Background sky levels were determined with
a first or second order fit in IRAF and subtracted from each object
spectrum.  The spectra for 50 separate regions of DIG were identified
in a similar manner, using apparent morphology and an upper cutoff in
emission measure; all DIG extractions have a surface brightness $\le$
100 pc cm$^{-6}$.  The DIG spectrum apertures vary from 10 pixels up
to 60 pixels.

Many of the large objects in our sample show significant internal
structure.  We have extracted separate spectra for each of these
structures using the H$\alpha$ emission variations as indicators of
boundaries.  The first spectrum extracted for each nebula contains
light from across the entire object in the slit.  Subapertures were
identified and labelled from left to right across the slit independent
of position angle.  All the apertures and subapertures are labelled in
Fig 1a and Fig 1b.  Each structure was classified based on the apparent
morphology and light profile in the H$\alpha$ images.  We have adopted
the classification scheme for the HII regions as published in
\markcite{wb92}WB92 with only a few slight variations described
further in the Appendix.

Emission-line fluxes were determined by fitting Gaussians to the line
profiles with the IRAF routine SPLOT.  Uncertainties were estimated by
measuring the rms noise on either side of the emission features with
an additional 3\% flux calibration error based on the spectra of the
standard stars.  The fluxes of the emission-lines measured in the
overlap region between the blue and red spectra (4000-6400$\rm\AA$)
were averaged together.  The lines included in this overlap region are
H$\gamma$, H$\beta$, [OIII]4959$\rm\AA$, 5007$\rm\AA$, and
[OI]6300$\rm\AA$.

\section{Extinction Corrections}

The emission lines of HII regions and SNR were corrected for
interstellar reddening using the Balmer decrement derived from the
observed H$\alpha$ to H$\beta$ line flux ratio for each spectrum
assuming a standard Galactic extinction curve (\markcite{sm79a}Savage
\& Mathis 1979).

The extinction in the faint extended DIG regions was addressed
carefully by \markcite{gwb97}Greenawalt, Walterbos \& Braun (1997).
They concluded that assuming a well-mixed dust layer rather than a
foreground dust layer produced a difference of less than 10\% in the
line flux ratios for the average observed Balmer decrement, and we
will use the foreground screen model here for the DIG as well as for
the HII regions and SNRs. We have assumed the same reddening
correction law for the DIG and the HII regions.  The intrinsic value
of H$\alpha$/H$\beta$ was assumed to be 2.86 for the HII regions and
DIG, based on Case B recombination and a gas T$_{e}$ $\approx$ 10,000K
(\markcite{o89}Osterbrock 1989).  The SNR H$\alpha$/H$\beta$ was taken
to be 3.00, the collisional value consistent with shock models of
\markcite{r79}Raymond (1979) and \markcite{sm79b}Shull \& McKee (1979).

The extinction and other properties of the nebulae are presented in
Table 2 and Table 3 for HII regions and SNRs, respectively. Column 3
lists the galactocentric radius for each nebula published in
\markcite{wb92}WB92. This radial distance is based on the spiral arm
most likely associated with the object as determined by the
kinematical model of \markcite{b91}Braun (1991).  Columns (4,5) list
the observed H$\alpha$/H$\beta$ line ratio and the extinction in
magnitudes, A(V). These parameters were used to correct each
individual spectrum for interstellar reddening.  

The dispersion of A(V) within an HII region can provide important
information about the distribution of dust and gas within the nebula.
It has been shown that the density within a single HII region is not
homogeneous (e.g. \markcite{k84}Kennicutt 1984). M31 is close enough
that our spectra can be used to investigate variations in A(V) across
large HII region complexes. By analyzing emission from a cross-cut of
an entire HII complex, we are no longer biased by selecting only the
bright, hence possibly less extincted sections of the HII region. We
can thus investigate if single extinction values derived for more
distant extra-galactic HII regions could be in error due to
significant extinction variations across the complexes.

In a recent study of HII regions in M33 and M101,
\markcite{pg97}Petersen \& Gammelgaard (1997) found that the variation 
of A(V) within large HII region complexes was typically 0.3 magnitudes
or less and that the extinction values usually increased toward the
edges of the nebulae.  We conducted a similar exercise for the largest
HII regions in our sample.  Fig 2 shows the A(V) variation within HII
regions complexes with more than 3 aperture extractions.  The
extinction obtained from each subaperture is plotted along with the
H$\alpha$ profile of the entire complex within the slit.

We note several points. Not surprisingly, the extinction measured for
the brightest region of a complex tends to be close to the derived
average extinction for that complex, a consequence of luminosity
weighting that is occuring in deriving an average extinction. However,
the brightest regions in a complex are not necessarily the ones with
the least extinction (e.g. K703, K525). This is important, because if
they were, then the derived average extinctions for complexes would be
systematically underestimated as a result of this luminosity
weighting. Typical variations in extinction across a complex amount to
$\pm 0.5$ mag. While this is not negligible, this variation is less
than that observed in the HII region population as a whole (panel g in
Fig 2). There is a faint region in the complex K703 which apparently
has significantly higher extinction than the rest of the complex (by 2
mag). In the present small sample, however, it is the only such region
and it is therefore not clear how common such variations are.

The peak in the distribution of A(V) (Fig 2g) for our sample is 0.8
magnitudes. The observed average foreground extinction toward M31 is
about 0.25 V mag (\markcite{bh84}Burstein \& Heiles 1984,
\markcite{ws87}Walterbos \& Schwering 1987).  The excess must be due to
internal extinction in the disk of M31. The peak in our distribution
is only slightly lower than the peak of the whole
\markcite{wb92}WB92 catalog of 958 nebulae, which was near 1.0
mag.  The extinction corrections in \markcite{wb92}WB92 were
determined by estimating of the amount of atomic hydrogen gas present
in front of each object.  Our extinction values, based on Balmer
decrements obtained from spectra, are much more accurate. While there
is overall agreement in the measured range of extinctions derived here
and in \markcite{wb92}WB92, the agreement for individual objects is
poor, which is not surprising given that the \markcite{wb92}WB92
method only provides a statistical estimate of the extinctions.  We
find no clear trends in A(V) with observed H$\alpha$ intensity, but we
do observe the highest measured extinctions in the annulus of vigorous
star-formation between 8 to 12 kpc, where the gas distributions also
peak (e.g. \markcite{w88}Walterbos 1988).

\section{HII Regions}

We have spectra for 46 separate HII regions from the
\markcite{wb92}WB92 catalog of the NE half of M31; this includes two
spectroscopically rejected SNR candidates which have been reclassified
as faint HII regions.  The spatial distribution of these nebulae is
concentrated in the high star-formation 10 kpc spiral arm, with the
fewest number of objects observed at the inner 5 kpc ring.  This makes
the study of radial trends, including the determination of the
chemical abundance gradient more uncertain.  However, we have observed
some correlations between line ratios and position in the galaxy as
well as spectral differences between nebulae of different
morphologies.  These are presented in \S4.2, while the next section
discusses the morphological classification of the HII regions.

\subsection{H$\alpha$ Morphology}

The general properties of our HII region sample are shown in Fig 3.
Average emission measures for each object were determined from the
reddening corrected total H$\alpha$ fluxes for each nebula using the
spectroscopically derived A(V) and the sizes given in
\markcite{WB92}WB92.  The nebular diameters are the average of
D$_{major}$ and D$_{minor}$ quoted in \markcite{WB92}WB92. The average
electron density, n$_{e,rms}$, is derived from the corrected H$\alpha$
emission measure assuming unit filling factor (E.M. = $\int$
n$_{e}^{2}$ dl).  It was also assumed that the HII regions are
spherically symmetric on average such that the path length through the
nebula along the line of sight is the same as the apparent linear
size.  It has been noted (\markcite{K84}Kennicutt 1984) that these are
unrealistic assumptions; we merely want to demonstrate the range in
properties in comparison to other HII region samples.  The dotted
lines in this plot represent lines of constant emission measure.

The point types used in Fig 3 distinguish between the morphological
classifications given in the \markcite{wb92}WB92 catalog:
center-brightened compact (shown as triangles), diffuse (five-pointed
stars) and rings (open circles).  We have also included the large HII
region complexes in these plots, shown as open squares.  The plot
shows that a rather smooth trend exists from the lower surface
brightness extended objects to the compact sources.  The distinction
between the compact sources and the extended objects is much greater
than between the diffuse, ring and HII region complexes.  This
separation also appears in the spectral properties of the objects, as
is shown in the next section.

\subsection{HII Region Emission-Lines and Nebular Conditions}

The most prominent HII region emission-line fluxes corrected for
interstellar reddening are presented in Table 4 in the form of line
ratios with respect to H$\beta$ or H$\alpha$.  Columns (1,2) list the
name of the object as designated in the \markcite{wb92}WB92 catalog
plus cross-references to \markcite{pavcmms78}Pellet \etal (1978)
objects, and the aperture sequence label.  The other columns list the
extraction aperture width in arcseconds (column 3), the H$\alpha$
emission measure for the region within the spectrum corrected for
interstellar reddening (column 4), and corrected line flux ratios:
[OII]3727$\rm\AA$/H$\beta$ (column 5); [OIII]5007$\rm\AA$/H$\beta$
(column 6); [NII]6548,83$\rm\AA$/H$\alpha$ (column 7); and
[SII]6717,31$\rm\AA$/H$\alpha$ (column 8).  The assigned morphology is
listed in the final column of Table 4.  We have used a similar
classification for the subapertures as that used for the discrete
nebulae in \markcite{wb92}WB92 with a few modifications; see the
Appendix for a complete description. Faint emission lines rarely found
in M31 HII regions were also detected in a few of our objects.  These
features include the [NeIII] forbidden lines and He I recombination
lines.  The extinction corrected fluxes for these lines are listed in
Table 5.

Electron temperatures and densities were derived from the appropriate
optical emission line ratios using the Lick Observatory FIVELEV code
(\markcite{ddh87}De Robertis, Dufour, \& Hunt 1987) and recently
updated atomic data.  Columns (6,7) in Table 2 list the density
sensitive [SII] line ratio and the derived n$_{e}$ (cm$^{-3}$) for the
HII regions.  Most of our objects are in the low density limit and
only two objects had marginally detectable [OIII]4363, required to
determine T$_{e}$ (\markcite{o89}Osterbrock 1989).  The electron
temperatures were estimated at T$_{e}$ $\sim$ 16,500K $\pm$ 4,000K for
K87 and T$_{e}$ $\sim$ 9,000K $\pm$ 1,000K for K932.  A canonical HII
region temperature of 10$^4$K was assumed for all other objects.

While the HII regions show emission-line ratios within the range
predicted by photoionization models, we do observe some spatial
variations in line flux even within the limited galactocentric radius
sampled by our data.  These variations are likely due to the chemical
abundance gradient in the disk of the galaxy but may also reflect
differences in the ionization level and excitation between nebulae at
different radii.

Due to the large inclination of the disk of M31, the galactocentric
radii of most of our objects are uncertain. In order to study the
effects of uncertain position in the galaxy, we have considered three
radial distance determinations in studying the spectral variations.
The first and least accurate distance (D$_1$) is determined by
assuming that the disk of M31 is an infinitely thin disk projected an
inclination angle of 77$^{\circ}$; the radial distances are based on
the x-y positions of the objects in this coordinate system.  The M31
kinematical model of \markcite{b91}Braun (1991) provides two possible
distance measurements (D$_2$, D$_3$), based on fitting spiral arms to
the HI kinematics; \markcite{wb92}WB92 lists these values as the
position of the first and second most likely spiral arm associated
with each source.

As is apparent in all the radial plots that follow, line ratio trends
with radius are, with some exceptions, marginal or not seen at
all. The plots show a lot of scatter even at a single radius.  While
all the radial plots were created using each of the possible distance
measurements, we have chosen to show only the plots based on D$_2$
which show radial gradients with the least scatter.  Undoubtedly, some
of the scatter in these plots is still likely due to the uncertainty
in the radial distance. However, much of the scatter is real and
presents an important problem in the interpretation of radial
gradients in emission-line ratios, as has been pointed out by
\markcite{bl97}Blair \& Long (1997). For example, the scatter may be
caused by internal variations in temperature, excitation or
density. Most of the data points are based on small sections of
nebulae that are already modest in luminosity to begin with. If the
HII regions can be ionized by just a few massive stars, the presence
or absence of just one bright O star can significantly affect the
spectrum. These effects are discussed more in Section~6. A line was
fit to the data points in each plot, using a weighted least squares
method. The fits are shown only if the resulting slope differed
significantly from 0, and if the correlation coefficient exceeded 0.5.

The first set of radial diagrams shows the variation of three oxygen
line ratios: [OIII]5007$\rm\AA$/H$\beta$ (Fig 4a-d),
[OII]3727$\rm\AA$/[OIII]5007$\rm\AA$ (Fig 4e-h), and R$_{23}$ (Fig
4i-l) defined by \markcite{ep84}Edmunds \& Pagel (1984) as the ratio
of the sum of [OIII]$\lambda$5007$\rm\AA$,
[OIII]$\lambda$4959$\rm\AA$, and [OII]$\lambda$3272$\rm\AA$ with
respect to H$\beta$.  The size difference in the symbols used
indicates if the data are from a discrete object (large size) or a
subaperture of an extended object or HII region complex (small size).
This distinction is almost equivalent to a luminosity coding of the
data points.

The line ratio [OIII]5007$\rm\AA$/H$\beta$, the excitation parameter
($\eta$), essentially measures the hardness of the radiation field
within the ionized nebula or equivalently, the temperature of the
central OB star(s).  This line ratio is also sensitive to chemical
composition (\markcite{s71}Searle 1971).  An outward radial increase
in [OIII]/H$\beta$ has been known to exist in disk galaxies for some
time (\markcite{a42}Aller 1942 and
\markcite{s71}Searle 1971) and is commonly used as evidence for an
oxygen abundance gradient in disk galaxies (\markcite{sdh92}Scowen,
Dufour, \& Hester 1992; \markcite{zhe90}Zaritsky, Hill, \& Elston
1990).  However, a strict empirical correlation between chemical
abundance and the observed [OIII]5007$\rm\AA$/H$\beta$ line ratio is
debatable since the line ratio only measures one ionization state of
oxygen (\markcite{ep84}Edmunds \& Pagel 1984; \markcite{z92}Zaritsky
1992). For M31, the expected increase in excitation with
galactocentric distance shows up only in the compact HII regions. The
diffuse nebulae, ring segments and DIG do not reproduce this trend and
appear only to scatter in the range covered by the compact sources. 

The high surface-brightness compact nebulae show a decrease in the
line ratio [OII]/[OIII] with increasing distance from the center, as
observed in HII regions in other galaxies.  The extended objects and
DIG again appear to show mostly scatter.  While the [OII]/[OIII] ratio
gives some information about the level of ionization in a nebulae, it
is also affected by metallicity.  To properly study the ionization of
HII regions, one would need to sample the multiple ionization states
of another element, like sulfur, but this requires spectral coverage
to the infrared ([SIII]9069$\rm\AA$ and [SIII]9532$\rm\AA$).

The R$_{23}$ parameter has been suggested as a good probe of oxygen
abundance in cases where the electron temperature, T$_e$, cannot be
determined directly from observations (\markcite{pebcs79}Pagel \etal
1979).  Several calibrations of this line ratio with oxygen abundance
have been published (\markcite{pes80}Pagel, Edmunds \& Smith 1980;
\markcite{ep84}Edmunds \& Pagel 1984; \markcite{v89}Vilchez 1989)
including the use of a weighted sum of the [OII] and [OIII] lines
(\markcite{bddb82}Binette \etal 1982).  Radial trends in the R$_{23}$
ratio have been observed and studied extensively in many nearby
galaxies (\markcite{e86}Evans 1986; \markcite{hh95}Henry \& Howard
1992; \markcite{gsssd97}Garnett \etal 1997; \markcite{kg96}Kennicutt
\& Garnett 1996).  We find a similar trend in our M31 data, although it 
again only appears in the center-brightened nebulae.

We studied the luminosity dependence of the oxygen emission-line
ratios with position in the galaxy to see if the lack of radial trends
and the large amount of scatter at any given radius could be explained
by differences in the number of ionizing stars within the nebulae.
Encoding the data points with the integrated luminosity of the nebula
from which the aperture was extracted indeed strengthens the radial
trends in the compact objects but has little effect on the scatter
observed in the other objects.

The radial plots of [NII]6583$\rm\AA$ and [SII]6716,31$\rm\AA$
relative to H$\alpha$ are shown in Fig 5.  There appears to be a
decreasing trend towards the outer disk of M31 in the [NII]/H$\alpha$
ratio in the lower surface brightness objects only. We will come back
to this in the discussion of the SNR spectra (\S 5.2 and \S 5.3). For
the bright compact HII regions, no such trend is observed, possibly
because a significant fraction of the nitrogen may be doubly ionized
there.

The radial distribution of H$\alpha$ surface brightness of the object
sections for which we have obtained spectra is shown in Fig 5i-l.
This figure shows that the observed emission-line ratio trends
discussed above are not caused by a correlation between H$\alpha$
surface brightness and radial distance.  We believe the observed
trends in oxygen and nitrogen are the consequence of the chemical
abundance gradient in M31 (see \S 5.3).

\section{Supernova Remnants}

\subsection{Confirmation of Candidates}

One of the original goals of this project was to spectroscopically
confirm the supernova remnant candidates from the \markcite{bw93}BW93
catalog.  This catalog consists of 52 SNR candidates which were
identified by high [SII]/H$\alpha$ line ratios ([SII]/H$\alpha$ $>$
0.45) in the narrow-band images published in \markcite{wb92}WB92.
Several of these candidates coincide with SNR candidate objects in
other catalogs such as \markcite{bkc81}Blair, Kirshner, \& Chevalier
(1981, hereafter BKC81) and \markcite{mpvlshpt95}Magnier \etal (1995).
A recent attempt has been made to confirm SNR candidates using the
ROSAT HRI (\markcite{mppvl97}Magnier \etal 1997).  Less than 10\% of
the optically identified candidates were detected in the X-ray,
however.  This can be explained as the result of a very tenuous
interstellar medium in the vicinity of SNR in M31, with typical
densities less than 0.1 cm$^{-3}$.  Thus, it is quite difficult to
confirm the candidates using X-ray emission alone and optical
spectroscopy continues to be the necessary and preferred method of
identifying these objects in M31.

Table 6 lists the spectral line ratios for all SNR candidate spectra.
Included in this table are the object name plus cross-reference to
BKC81 and Blair, Kirshner \& Chevalier (1982, hereafter BKC82) and
aperture sequence label in Columns (1,2), aperture width in arcseconds
in Column (3), and reddening corrected H$\alpha$ emission measure in
Column (4).  Columns (5-9) list the reddening corrected flux ratios of
[OII]3727$\rm\AA$ and [OIII]5007$\rm\AA$ to H$\beta$ and
[OI]6300$\rm\AA$, [NII]6548,83$\rm\AA$ and [SII]6716,31$\rm\AA$ to
H$\alpha$.

To address the candidacy of our objects, we first compared the
observed spectroscopic [SII]/H$\alpha$ line ratio with the ratio
obtained from the narrow-band line images for the exact object section
captured in each spectrum.  In all but a few rare cases, there is a
good agreement between the spectral ratio and image ratio.  Bad pixels
in the line images or in the 2D spectra images affected the agreement
in a few cases; poor signal-to-noise also played a factor for a few
candidates.

The morphology of the candidates studied here include discrete
ring-like structures, faint diffuse nebulae and compact sources
embedded in nebular complexes (indicated by an 'A' in the object
name).  Most of the SNR extractions show enhanced [OII]3727$\rm\AA$
and [OIII]5007$\rm\AA$ emission, and a few of the objects show a
bright [OI]6300$\rm\AA$ feature in their spectra.  These factors taken
together help in the confirmation of these candidates.  In several
cases, the gas surrounding embedded compact SNR candidates does not
show enhanced [SII] emission relative to H$\alpha$.  Two SNR
candidates (K310 and K446) were rejected as supernova remnants due to
low [SII]/H$\alpha$ and lack of bright [OIII] emission.  It is
suspected that these objects are in fact HII regions.  SNR candidates
of particular interest are described in the Appendix.

\subsection{SNR Line Ratios and Nebular Conditions}

The brightness of optical emission lines in shocked gas depends on
multiple factors: the electron density and electron temperature in the
post-shock region which are related to the velocity of the shock
front; the density of the medium into which the shock is propagating;
and the chemical abundance of elements of the swept up material, among
other things.  The modelling of these effects on the optical spectra
of SNR has been carried out by \markcite{dmf77}Dopita, Mathewson \&
Ford (1977), \markcite{sm79b}Shull \& McKee (1979), and
\markcite{r79}Raymond (1979).  Through a series of diagnostic diagrams
based on the ratios of optical emission lines, it has been shown that
the variations in the spectra of SNRs are due primarily to the
chemical abundance effects and only minimally to the shock conditions
for shock velocites near and above 100 km s$^{-1}$
(\markcite{ddbdb84}Dopita \etal 1984a).  However, the modelling of
interstellar shocks is not straightforward due to the large number of
variables involved and the theoretical models do not always agree with
observations.  There are many complications due to the depletion of
volatiles onto grains, the destruction of grains and even magnetic
field effects.  However, there are several basic line ratio
diagnostics which can help probe the physical conditions of the
shocks.

\markcite{bkc82}BKC82 have used the variation in the observed line 
flux ratios of SNR to derive the abundance gradient in the disk of
M31.  The spatial distribution of our confirmed candidates is limited
to the outer regions of the disk and we have no objects at
galactocentric radii smaller than 8kpc.  Thus we cannot use our data
to directly measure the abundance gradient in M31.  However, the
general radial trends found by \markcite{bkc82}BKC82 are fairly well
reproduced with our new data.  For comparison, we have plotted the
\markcite{bkc82}BKC82 data points (in open squares) as well as our
confirmed SNR data points (solid squares) in Fig 6.  The agreement in
the slopes and also in the scatter is good given the faintness of most
of our remnants relative to those studied by \markcite{bkc82}BKC82.

Fig 6a shows the radial variation of the excitation line ratio
[OIII]5007$\rm\AA$/H$\beta$ for SNRs.  Contrary to the trend found in
the HII region spectra, the SNR spectra do not show a clear increase
in excitation with distance from the center, though there is
significant scatter in the ratio at any given position.  Since the
[OIII] emission arises from a region very close to the shock front,
two competing factors govern the brightness of the [OIII] lines: the
postshock temperature (equivalent to shock velocity) and the oxygen
abundance (\markcite{d77}Dopita 1977).  The ionization line ratio,
[OIII]5007$\rm\AA$/[OII]3727$\rm\AA$, on the other hand, has been
shown to be more sensitive to the postshock conditions and less
sensitive to metallicity (\markcite{d77}Dopita 1977).  Our data in Fig
6b show fair agreement with the BKC points with more scatter, likely
due to measurement error.  These two emission line ratios do not
likely indicate a chemical abundance gradient, but reflect a gradient
in the SNR physical conditions across the galaxy.

Our SNR data also show the correlation between [OII] and [OIII]
emission found by \markcite{bks82}BKC82 (Fig 6c).  The curves plotted
here are shock model predictions for metallicity variations (thick
solid curve), shock front velocity variation (dash-dot curve) and
shock velocity variation with the pre-ionation of the medium (dash
curve) taken from \markcite{dbdb84}Dopita \etal (1984).  The behavior
of the two oxygen line ratios can be reproduced by varying the
abundance of oxygen relative to hydrogen (by number) but not by
varying the shock conditions.  This indicates that the group of SNRs
studied here do show a wide range of oxygen abundance.

The [NII]/H$\alpha$ trend shown in Fig 6d shows the cleanest radial
trend.  The [NII] lines are not strongly affected by shock temperature
since they originate in the large recombination region behind the
shock front.  These lines are also not affected by collisional
de-excitation and so are relatively insensitive to electron density.
This trend must be a direct result of the abundance gradient in M31.
The [SII]/H$\alpha$ gradient (Fig 6e), may also primarily show an
abundance effect.

The physical conditions derived from the optical spectra for our
sample of SNRs are listed in Table 3 (density sensitive [SII] line
ratio and n$_{e}$ in Columns (6,7) respectively) and Table 7 (T$_e$).
Electron densities were determined for those regions with good [SII]
doublet line detections ([SII]6716$\rm\AA$/[SII]6731$\rm\AA$ within
the theoretical limits) using the Lick Observatory FIVEL program with
the assumption that the recombination zone from which the sulfur
doublet lines arise has a temperature of 10$^{4}$K.  In most cases,
the SNRs are in the low density limit.  This has been recently
confirmed by X-ray data which show that upper limits to the electron
density in the interstellar medium in the vicinity of SNR candidates
in M31 is about 0.1 cm$^{-3}$ (\markcite{mppvl97}Magnier \etal 1997).

Unlike HII regions, the conditions in the postshock gas of SNR are
such that the high excitation emission-line [OIII]4363$\rm\AA$ is
easier to detect.  This line was detected in a few of our SNR
candidates spectra, allowing the determination of T$_{e}$.  Using the
formalism of \markcite{kace76}Kaler \etal (1976) and the
[OIII]4959$\rm\AA$ + 5007$\rm\AA$/[OIII]4363$\rm\AA$ emission-line
ratio, we have determined the temperatures listed Table 7.

In Fig 6f, we plot the electron temperatures derived for the SNR with
available [OIII]4363$\rm\AA$ measurements as a function of radial
distance.  Our data points are in fair agreement with the
\markcite{bkc82}BKC82 points at radii near 10kpc.  There are too few
points in the outer regions of the galaxy to see a spatial correlation
in the post-shock temperature in the galaxy.  Such a correlation would
explain the slight increase of the [OIII]/H$\beta$ and [OIII]/[OII]
line ratios towards the center of the galaxy shown in Fig 6a and 6b.

\subsection{Summary of abundance trends in HII regions and SNRs}

Despite our limited radial sampling of the disk of M31, we have
detected radial trends in some of the emission lines for both HII
regions and SNRs which likely reflect the abundance gradient in
M31. We compare our results, which concern mostly low-luminosity
objects, with those of \markcite{bkc81}BKC81 and \markcite{bkc82}BKC82
who concentrated on the brightest objects and covered a larger radial
range.  \markcite{dbdb84}Dopita \etal (1984) reanalyzed the Blair
\etal\ data with improved shock ionization models, to resolve a
discrepancy in the derived oxygen abundances for HII regions and SNRs.

The oxygen line ratios [OIII]/H$\beta$, [OII]/[OIII], and R$_{23}$
observed in the center-brightened HII regions (Fig 4a, 4e, 4i) show
the expected correlation with radial position within the galaxy.
These line ratios are all consistent with increasing metallicity
towards the center of the galaxy.  The gradient we can infer from the
R$_{23}$ parameter is -0.06 $\pm$ 0.03 dex kpc$^{-1}$, in fair
agreement with the gradient derived from HII region spectra by
\markcite{bkc82}BKC82 and the gradient based on shock-ionization
modeling of SNR line ratios (\markcite{dbdb84}Dopita \etal 1984),
-0.05 $\pm$ 0.02 dex kpc${-1}$.  Peculiarly, none of the other types
of HII regions show a strong radial gradient in the oxygen emission
lines.

The correlation between the [OII] and [OIII] emission from the SNRs
(Fig 6c) indicates that these objects also exhibit metallicity
differences (\markcite{dbdb84}Dopita \etal 1984).  The radial
distribution of [OII]/H$\beta$ and [OIII]/H$\beta$ observed in our
sample of SNRs is in good agreement with the line ratios found in
\markcite{bkc82}BKC82 sample of SNRs.  This further confirms the
oxygen abundance gradient determined from the proper shock-ionization
treatment of the \markcite{bkc82}BKC82 SNR emission lines by
\markcite{dbdb84}Dopita \etal (1984).

The center-brightened HII regions do not show a radial gradient in
[NII]/H$\alpha$ over the range 5 to 15 kpc (see Fig 5), consistent
with \markcite{bkc82}BKC82, who mainly show higher ratios for this
line within 5 kpc from the center. However, the [NII]/H$\alpha$ ratios
are correlated with radial distance in the case of the ring-like
nebulae which show a gradient in [NII]/H$\alpha$ (Fig 5) of -0.02
$\pm$ 0.01 dex kpc $^{-1}$; the diffuse nebulae and the DIG also show
this trend, but with more scatter. The photoionization models for
these low-excitation regions (e.g. Domg\"orgen \& Mathis 1994) indeed
show the line ratio of [NII]/H$\alpha$ to correlate with the nitrogen
abundance. These authors note that the forbidden line ratios depend on
electron temperature and abundance in complicated ways, but their
calculations show a monotonic pattern for [NII]/H$\alpha$, while the
[SII]/H$\alpha$ model ratio trends in \markcite{dm94}Domg\"orgen \&
Mathis (1994) with abundance seem to be more complex.  This may
explain why our data do not show a radial trend in [SII]/H$\alpha$
ratios.  In comparison, the SNRs show a steeper gradient of -0.04
$\pm$ 0.01 dex kpc $^{-1}$ in the [NII]/H$\alpha$ line ratio (Fig 6d).
This discrepancy, although not explicitly noted by
\markcite{bkc82}BKC82, can be reproduced using their HII region and
SNR data.  Both the average value of the [NII]/H$\alpha$ line ratio
and the radial gradient of this line seen in the SNR and the HII
regions differ significantly.  However, the methods used to determine
the nitrogen abundance in the photoionized nebulae and the
shock-ionized nebulae are distinct, so despite differences in the line
flux ratios, the nitrogen abundances derived from both types of
objects are consistent with each other (\markcite{bkc82}BKC82).  Our
ring-like and diffuse nebulae, and SNR line ratios are consistent with
the HII regions and SNRs in
\markcite{bkc82}BKC82, so we confirm the abundance gradient published
previously for the nitrogen in M31, approximately -0.07 dex
kpc$^{-1}$.

\section{Discussion}

The optical spectra of HII Regions, SNRs, and regions of DIG make it
possible to compare the emission-line properties of these different
classes of objects.  Diagnostic line-ratio diagrams are a good way to
constrain the ionization and excitation properties of various phases
of the ISM.  This type of spectral analysis is especially important in
the case of the DIG where the ionization mechanism has not been well
constrained.  We consider observed line ratio trends with H$\alpha$
surface brightness, present diagnostic diagrams which separate stellar
photoionized from shock-ionized nebulae, and discuss how a smooth
transition between HII regions and DIG is indicated.

\subsection{Spectral Line Transitions: HII Regions and DIG}

As noted by several authors (\markcite{m97}Martin 1997;
\markcite{whl}Wang, Heckman \& Lehnert 1997; \markcite{gwb97}Greenawalt, 
Walterbos \& Braun 1997), there is a strong correlation between the
[SII]/H$\alpha$ line flux ratio and the H$\alpha$ surface brightness
for HII regions and DIG.  Is such a trend present for other forbidden
lines as well?

The four panels of Fig 7 show the line ratio variations with H$\alpha$
emission measure in HII regions and DIG for [OIII]/H$\beta$,
[SII]/H$\alpha$, [NII]/H$\alpha$, and [OII]/H$\beta$. Errorbars have
only been included for the DIG to reduce clutter; errorbars for the
HII region data are typically much smaller due to higher
signal-to-noise.

No systematic variation is observed in the [OIII] emission with
changing surface brightness in either the HII regions or in the DIG.
In fact, the DIG [OIII] emission is comparable to that of HII regions.
In the \markcite{whl97}Wang, Heckman \& Lehnert (1997) scenario which
proposes the existence of two types of DIG, this suggests that we may
be mostly seeing diffuse gas which resides in the plane of the galaxy
and is associated with star-formation.  Strong [OIII] emission would
be consistent with a second type of DIG, a disturbed phase which may
exist higher above the plane of disks galaxies and may be
shock-ionized (\markcite{whl97}Wang, Heckman \& Lehnert, 1997;
\markcite{r98}Rand 1998).  Of course, the relatively edge-on
orientation of M31 might hide the faint emission from this gas far
above the midplane of the galaxy, but there are other reasons to
believe that M31's DIG layer is not very thick
(\markcite{wb94}Walterbos \& Braun 1994).  We have already mentioned
that the [OIII] emission from the center-brightened HII regions varies
as a function of position in the galaxy due to the metallicity
gradient in M31. This increase is not seen in the DIG surrounding
these HII regions (see \S 3.3), which may explain why there is less
scatter in this ratio for the DIG than for HII regions.

The strongest correlation observed to date is the relative increase in
the [SII] emission with decreasing H$\alpha$ surface brightness (Fig
7b).  The [SII]/H$\alpha$ ratio has commonly been used as one of the
defining characteristics separating HII regions and DIG, the first
being the low surface brightness in H$\alpha$ for the DIG, coupled
with diffuse morphology.  Most HII regions have [SII]/H$\alpha$ ratios
below 0.3 while the DIG has ratios which reach up to 1.0.  The
interesting point about this plot is the positioning of the diffuse
and ring nebulae relative to the center-brightened sources and the
DIG.  These points seem to fill the gap between the two brightness
extremes creating a smooth sequence in the [SII]/H$\alpha$ ratio. The
extended nebulae may be ionized by a more diffuse radiation field than
the center-brightened sources.  Thus the smooth trend observed in this
plot is likely the result of a smoothly decreasing ionization
parameter from the compact sources down to the most diffuse ionized
gas.  This trend is reproduced in the fainter [NII] emission (Fig 7c)
albeit with much more scatter than the [SII] lines.  In this case,
abundance effects may play a role in the scatter.

As is the case for [OIII], the [OII] emission (Fig 7d) is not strongly
correlated with H$\alpha$ emission measure in the HII regions or the
DIG.  One might have naively expected to see the same increasing trend
with decreasing H$\alpha$ emission measure in all emission lines
arising from singly-ionized species if the increase in flux is due to
a decreasing ionization parameter.  However, the photoionization
models of \markcite{dm94}Domg\"orgen \& Mathis (1994) predict that
only the [NII] and [SII] flux will increase as the ionization
parameter decreases.  The [OII] emission is nearly independent of the
ionization parameter and thus remains fairly constant in the HII
regions and in the DIG.

\subsection{Diagnostic Diagrams: HII Regions, SNR and DIG}

Over the last two decades, theoretical modeling of the emission line
spectrum of ionized gas has proven quite useful in determining various
characteristics of the gas.  Line ratio diagnostics of excitation
conditions were investigated by \markcite{bpt81}Baldwin, Phillips \&
Terlevich (1981, hereafter BPT81), resulting in a set of diagrams
which separate photoionized gas from shock-ionized gas.
\markcite{ed85}Evans \& Dopita (1985) developed a set of solar 
metallicity HII region grids on these diagrams to study the effects of
varying the stellar ionization temperature T$_{ion}$, the mean
ionization parameter {\it$\bar{Q}$}(H), and the element-averaged
metallicity $\bar{Z}$.  These grids are based on photoionization
models for steady-state spherically symmetric nebulae with unit
filling factor and a centrally located OB association, all assumptions
which do not necessarily hold in real HII regions.  However,
observational data on HII regions fit quite nicely into these grids
showing relatively good agreement between known conditions and those
predicted by the models.  

Here we use recent ionization models which predict line ratios for
various ionization mechanisms, ionization parameters and nebular
conditions (eg. \markcite{sf90}Shields \& Filippenko 1990;
\markcite{s93}Sokolowski 1993; \markcite{dm94}Domg\"orgen \& Mathis
1994) on these diagnostic diagrams to study the differences in the
spectra of the different types of HII regions, DIG and SNRs.  Although
in all cases we have used reddening corrected line ratios, one of the
advantages of diagnostic diagrams is that most of the ratios used are
based on lines very near in wavelength so errors in the reddening
correction will not affect our results.

The first set of diagnostic diagrams compares the [OIII] and [NII]
emission to the [SII] emission in the HII regions, DIG and SNR.  The
group of data points clearly shifts towards higher [SII]/H$\alpha$ in
each consecutive panel (Fig 8a-d).  The center-brightened HII regions
fall completely within the photoionization model box (labelled as
``HII region'''') of \markcite{sf90}Shields \& Filippenko (1990) while
the diffuse and ring-like nebulae begin to scatter outside this box
towards shock ionization models (labelled as ``LINER'').  The DIG and
SNRs are found well within the box for shock-ionization with very few
exceptions.  Although the [SII] emission from the DIG is inconsistent
with the photoionization models of \markcite{sf90}Shields
\& Filippenko (1990), the predictions of \markcite{s93}Sokolowski (1993)
and of \markcite{dm94}Domg\"orgen \& Mathis (1994) which specifically
addressed photoionization by a dilute radiation field, do reproduce
the enhancement of the [SII] emission in the lower emission measure
gas.  The arrows on the model curves indicate the direction of
decreasing ionization parameter.

The [NII] and [SII] emission features can be used together to best
separate photoionization and shock-ionization mechanisms (Fig 8e-h).
All confirmed SNR in our data have by definition [SII]/H$\alpha$ $\ge$
0.45, so we have divided up each of the plots by a dashed line at
log([SII]/H$\alpha$) = -0.35.  From the plots, it is apparent that
there also seems to be a lower limit on the [NII] emission of the
shock-ionized nebulae near log([NII]/H$\alpha$) = -0.3.  We have
placed a dashed lined at this value to split up each plot into four
quadrants.  Only the SNR occupy the upper right corner while the HII
regions and DIG are found mostly outside of this quadrant.  The two
\markcite{dm94}Domg\"orgen \& Mathis (1994) models (``leaky'' and 
``composite'') are in fair agreement with the observed spectra of the
DIG, showing the increase in [NII] with a more rapid increase in the
[SII] emission as the ionization parameter decreases.

Two classic \markcite{bpt81}BPT81 excitation plots based on the
ionization line ratio [OII]/[OIII] are given in Fig 9.  The first set
of panels shows the correlation between [OIII]/H$\beta$ and
[OII]/[OIII].  The center-bright HII regions lie completely below the
upper envelope of photoionization models (\markcite{ed85}Evans \&
Dopita, 1985). The diffuse and ring nebulae show a bit more scatter
which places a few points above this curve.  The one point lying far
above this curve corresponds to the bright ring segment, K450.a which
shows extraordinarily high reddening due to a very weak H$\beta$ flux.
The DIG (Fig 9c) also lie mostly below this curve, suggesting the
oxygen line ratios are consistent with even the simplest
photoionization models. The \markcite{dm94}Domg\"orgen \& Mathis
(1994) models at the higher ionization parameter predict the [OIII]
emission fairly well.

The SNRs (Fig9d) cluster about the upper photoionization curve with
most points lying above the curve, as expected, although the number of
points lying below the theoretical curve is substantial.  The [OIII]
emission in a SNR is sensitive to both the shock velocity and chemical
abundance, as has already been pointed out, so the points below this
curve are not necessarily consistent with photoionization model
predictions.

Placing our data points in the [NII]6584/H$\alpha$ versus [OII]/[OIII]
plane (Fig 9e-h) also separates the two ionization mechanisms.  The
dashed-line boxes represent theoretical models for photoionization and
shock ionization based on \markcite{sf90}Shields \& Filippenko (1990).
As in previous plots we also show the models of
\markcite{dm94}Domg\"orgen \& Mathis (1994).  Although the
center-brightened nebulae show a wide range in the ionization ratio,
the [NII] emission remains fairly constant around the typical HII
region value of log([NII]/H$\alpha$)$\approx$-0.4.  The DIG, diffuse
and ring-like HII regions (in Fig 9b and 9c) show a similar
distribution consistent with photoionization with only a few points
scattering into the shock model regime.

However, the SNRs cluster about an average value of
log([NII]/H$\alpha$) $\approx$ -0.2, forcing the points to lie in the
shock-ionization box.  Again, there are numerous SNRs which lie below
the shock-ionization model box and there appears to be even more
overlap between the SNRs and the HII regions than in Fig 9a-d.  This
is most likely an abundance effect, since the [NII] lines in SNRs are
sensitive to chemical composition not the shock conditions.  The SNRs
which fall below log([NII]/H$\alpha$)=-0.2 still show the elevated
[SII] emission expected from shock conditions as is shown in Fig 8h.

Analysis of the emission-line diagnostic diagrams leads to the
following conclusions:

\begin{itemize}

\item{The enhanced [SII] emission from the DIG are the only forbidden
lines not clearly reproduced by simple photoionization models, that
would seem to be more consistent with shock-ionization.  However,
dilute photoionization models such as those presented by
\markcite{dm94}Domg\"orgen \& Mathis (1994), do explain the enhanced
[SII] emission while keeping the [NII], [OII], and [OIII] line
strengths near the values found in typical HII regions.  The optical
emission lines from the DIG observed thus far in M31 are consistent
with the low ionization parameter photoionization models as described
in \markcite{gwb}Greenawalt, Walterbos, \& Braun (1997).}

\item{The diagnostic diagrams presented here separate the 
spectroscopically confirmed SNRs from the photoionized nebulae fairly
well, the best case being that of the [NII]/H$\alpha$
vs. [SII]/H$\alpha$ plot (Fig 8).  The shock-ionized nebulae exist in
a very restricted quadrant of this plot with enhanced [SII] and [NII]
emission.  The DIG does not occupy this quadrant.}

\item{The line ratios for the diffuse HII regions and the ring nebulae 
are consistent with photoionization models and form a transition
between center-brightened HII regions and the DIG in terms of the
[SII]/H$\alpha$ ratio.}

\end{itemize}

\section{Summary}

We have obtained deep optical spectroscopic observations for 46 HII
regions (including 2 rejected SNRs), 16 confirmed SNRs, and numerous
regions of diffuse ionized gas in M31.  The majority of the HII
regions studied here are in the low density limit and have low enough
temperatures that direct determination of T$_{e}$ by the observation
of the [OIII] 4363$\rm\AA$ line is impossible.  This complicates the
determination of the abundance gradient in M31, although the observed
variations of the line ratios with radius indicate that a gradient
does exist.  The abundance gradient we determine for M31 based on the
R$_{23}$ parameter is -0.06 $\pm$ 0.03 dex kpc $^{-1}$, consistent
with all previously determined values.

This radial variation of line ratios does appear to depend on
morphology.  The higher surface brightness HII regions show spectra
which are dominated by local chemical abundance effects and thus can
be used to trace out the chemical gradient in the galaxy.  The spectra
of the diffuse and ring HII regions do not show radial line ratio
trends and appear to be dominated by ionization and excitation effects
rather than metallicity.

We have confirmed 16 of the 18 observed SNR candidates from the
\markcite{bw93}BW93 catalog.  The two remaining objects were 
rejected due to low [SII] emission.  These objects also do not show
enhanced [OIII] or [OI] emission in their spectra and were thus
reclassified as normal HII regions.  The radial trends in the line
ratios for the confirmed SNR agree with previously published results
(\markcite{bkc81}BKC81 and \markcite{bkc82}BKC82).

While the [SII] emission from the HII regions and DIG shows a strong
increase with decreasing H$\alpha$ emission measure, no other spectral
feature exhibits such behavior.  The smooth transition in
[SII]/H$\alpha$ seen from the center-brightened HII regions to the DIG
strongly suggests that the DIG is affected by a lower ionization
parameter, diluted radiation field.  The diffuse and ring-like HII
regions which exhibit strong ionization and excitation effects in
their optical spectra, represent transitional objects between the
bright HII regions and the faint DIG.

Using diagnostic diagrams which compare the [OIII], [NII], and [SII]
emission to the brightest Balmer lines, we have shown how most of the
emission features found in the DIG are consistent with even the
simplest photoionization models.  However, the [SII] emission from the
DIG cannot be explained by these simple models.  The increase in [SII]
emission with decrease in H$\alpha$ surface brightess we have observed
is well explained by the diluted radiation field photoionization
models of \markcite{dm94}Domg\"orgen \& Mathis (1994).  There is no
evidence for shock ionization in the DIG layer of M31.

\acknowledgments

We would like to acknowledge the insightful discussions with Don
Garnett on abundance gradient effects on emission-line ratios in
spiral galaxies, and Stacy McGaugh on empirical abundance calibrators.
We graciously thank Charles Hoopes for reading and commenting on the
manuscript during its development.  The comments from an anonymous
referee helped to clarify presentation of the results.  This work has
been supported by NSF grant AST-9617014 and a Cottrell Scholar
Award from Research Corporation to R.A.M.W.

\section{Appendix : Special Cases}

Of the 46 discrete HII regions observed we have identified 9 diffuse
nebulae, 16 center-bright nebulae, 12 extended complexes, and 9
discrete ring structures.  In a very few cases, the new classification
used is in disagreement with the \markcite{wb92}WB92 classification.
The subapertures obtained for structures within extended nebulae have
also been classified using the same nomenclature.  It is important to
note, in this case, that a morphological classification of ``ring''
can mean that the spectrum contains light across an entire object
which appears as an H$\alpha$ ring or can mean that the spectrum is
from a section of a ring; a distinction is made only in Fig 4 and Fig
5 where the point size indicates if the object is a discrete nebula or
a substructure within a nebula.

We have also obtained spectra of large complexes and other nebulae
within which are embedded SNR candidates from the \markcite{bw93}BW93
catalog.  Here we provide a set of brief descriptions for these
objects as well as for objects which have been re-classified or which
require a more detailed description of their morphology.  The central
positions for apertures are labelled in the finding charts, Fig 1a and
Fig 1b.

\noindent
\begin{bf}K87\end{bf} was classified as a center-brightened source 
by \markcite{wb92}WB92 although it consists of a bright compact source
(K87.b) centered within a faint diffuse shell of emission
approximately 150 pc in diameter.  It is unclear whether this diffuse
ring of gas is really associated with the bright knot, especially
since the spectral properties are quite different.  The line fluxes in
the spectrum of the west filament (K87.a) are included in Table 4; the
east filament had insufficient signal-to-noise to include in our
results.

\noindent
\begin{bf}K103.a\end{bf} is a filamentary structure which appears on 
the far northern edge of a confirmed supernova remnant, K103A.
Several spectra were extracted to study the complicated structures in
the large complex, K103.  The spectra of various knots and filaments
do show the elevated [SII]/H$\alpha$ ratio indicating shock-ionization
and confirm the nature of the embedded supernova remnant candidate.
The filament K103.a, however, has a very low [SII] ratio, suggesting
that it is actually not part of the supernova remnant.

\noindent
\begin{bf}K132\end{bf} is listed as a diffuse nebula in 
\markcite{wb92}WB92 but appears to be a bright compact source embedded 
within an amorphous diffuse nebula.  This knot is a center-brightened
25 pc HII region.  The diffuse gas surrounding this object is too
faint to obtain a good spectrum.

\noindent
\begin{bf}K310\end{bf} is one of two \markcite{bw93}BW93 supernova 
remnant candidates we have rejected due to the low [SII]/H$\alpha$
ratio observed in its spectrum.  It is a 55 pc, very diffuse and faint
(E.M. $\approx$ 90 pc cm$^{-6}$) amorphous nebula.  Though it is
fainter than our adopted DIG cutoff surface brightness, it is not
classified as DIG because it appears as a discrete nebula.

\noindent
\begin{bf}K446\end{bf} was also suspected to be a supernova
remnant due to the high [SII]/H$\alpha$ ratio obtained from line
images.  Our spectroscopy has ruled out this candidate however, which
shows a ratio of only 0.24.  The low ratio determined from the images
was affected by a few bad pixels.

\noindent
\begin{bf}K496\end{bf} is a large HII region which contains
an embedded SNR candidate, K496A (\markcite{bw93}BW93).  We did not
obtain a spectrum of the candidate however.  The section of the object
for which we have obtained a spectrum does not show the signature
enhanced [SII]/H$\alpha$ line ratio so we have included K496 with
the HII regions.

\noindent
\begin{bf}K525\end{bf} is a large HII region complex consisting of 
several diffuse structures, faint arcs, bright compact knots and
includes an embedded supernova remnant candidate, K525A
(\markcite{bw93}BW93) toward the center of the region.  All but the
central aperture which corresponds to the supernova remnant (K525.c),
  show [SII]/H$\alpha$ ratios typically found in photoionized nebulae.

\noindent
\begin{bf}K526\end{bf} The spectrum obtained for this HII region 
complex contains light from an embedded supernova remnant candidate
K526A, and a center-bright nebula with a diameter near 80 pc.  The
[SII]/H$\alpha$ ratio in two spectra (K526.b and K526.c) are enhanced,
confirming the embedded SNR; the other aperture, K526.a, which
corresponds to the compact source, shows [SII] emission typical of HII
regions.

\noindent
\begin{bf}K877\end{bf} is a large (d $\approx$ 200 pc) complex which 
contains an embedded Wolf-Rayet star (\#1201 in \markcite{mla93}MLA93)
near the center of a small ring shaped structure.

\noindent
\begin{bf}K932\end{bf} was classified as a single object in the 
\markcite{wb92}WB92 catalog, though closer inspection of the 
H$\alpha$ image reveals that it is actually a cluster of several very
bright compact sources and may include a small faint complete ring to
the north.  The first spectrum (K932.a) is that of a small knot near
the rim of the faint ring-like HII region.  The second (K932.b) is a
bright compact source with a diameter of about 40 pc.  We have listed
K932 as a complex because it does not appear to be a discrete nebula.

\newpage

\begin{deluxetable}{lccrcl}
\tablecolumns{5}
\scriptsize 
\tablewidth{0pc}
\tablenum{1}
\tablecaption{Slit Positions} 
\tablehead{
\colhead{Spectrum} 
& \multicolumn{2}{c}{Slit Center} 
& \colhead{PA\tablenotemark{1,2}} 
& \colhead{WB92} 
& \colhead{Target Objects}\\
\colhead{} 
& \colhead{RA(2000)} 
& \colhead{DEC(2000)} 
& \colhead{ } 
& \colhead{Field} 
& \colhead{ }  
}
\startdata
K59  & 0:43:01.0 & 41:38:28.5 & 156.9 & 15 & K59, K87, K82, K91, K92 \\
K86  & 0:43:00.0 & 41:36:54.5 & 135.4 & 15 & K68, K70, K76, K78, K81, K86\tablenotemark{3} \\
K103 & 0:43:21.0 & 41:44:05.4 & 129.6 & 14 & K132, K145, K103\tablenotemark{3,5} \\
K230 & 0:43:56.0 & 41:12:46.3 & 238.0 &  3 & K230\tablenotemark{3}, K244, K250, K252\tablenotemark{3} \\
K310 & 0:44:10.0 & 41:33:03.3 &  92.6 &  1 & K310\tablenotemark{4}, K314, K315, K316 \\
K327 & 0:44:17.0 & 41:18:30.3 & 119.6 &  4 & K330, K327\tablenotemark{4}, K343, K353 \\
K425 & 0:44:25.0 & 41:20:06.3 & 105.0 &  4 & K391, K403, K425\tablenotemark{4} \\
K434 & 0:44:34.1 & 41:51:48.2 & 180.9 & 12 & K531, K496\tablenotemark{3}, K480, K434, K442 \\
K446 & 0:44:28.0 & 41:21:46.3 & 147.5 &  5 & K414, K446\tablenotemark{3}, K450 \\
K506 & 0:44:39.0 & 41:25:32.2 & 185.0 &  6 & K447, K461, K490\tablenotemark{3}, K506\tablenotemark{3}, K536\tablenotemark{6}\\
K526 & 0:44:42.0 & 41:27:02.2 & 103.5 &  6 & K525\tablenotemark{3}, K526\tablenotemark{3}, K527\tablenotemark{3}, K536\tablenotemark{6}\\
K594 & 0:44:59.1 & 41:55:21.2 & 195.4 & 11 & K594\tablenotemark{4}, K653\tablenotemark{5} \\
K717 & 0:45:11.1 & 41:37:35.2 & 106.1 &  8 & K703, K722, K717\tablenotemark{4}\\
K856 & 0:45:37.1 & 41:55:06.1 & 207.5 & 10 & K772, K787, K838, K851, K856\tablenotemark{3}\\
K884 & 0:45:51.1 & 42:10:29.1 & 222.6 & 19 & K884\tablenotemark{4}, K877\tablenotemark{5} \\
K934 & 0:46:34.2 & 42:12:03.0 & 265.0 & 18 & K927, K931, K932, K934\tablenotemark{4} \\
\tablenotetext{1} {Slit position angles are listed in degrees.}
\tablenotetext{2} {Exposure time for each slit position is 30 minutes.}
\tablenotetext{3} {Object contains an embedded Braun \& Walterbos (1993) SNR candidate.}
\tablenotetext{4} {Object is listed as a Braun \& Walterbos (1993) SNR candidate.}
\tablenotetext{5} {Object contains a Wolf-Rayet star candidate.}
\tablenotetext{6} {Object was observed at two position angles.}
\enddata
\end{deluxetable} 

\newpage
\begin{deluxetable}{lcrrrrrrr}
\tablecolumns{9}
\scriptsize 
\tablewidth{0pc}
\tablenum{2}
\tablecaption{HII Region Nebular Conditions} 
\tablehead{
\colhead{Object\tablenotemark{1}} 
& \colhead{Aperture}
& \colhead{R$_g$\tablenotemark{2}} 
& \colhead{H$\alpha$/H$\beta$\tablenotemark{3}}
& \colhead{A$_{V}$\tablenotemark{4}} 
& \colhead{R$_{[SII]}$\tablenotemark{5}} 
& \colhead{n$_e$}  \\
\colhead{ }  
& \colhead{ }  
& \colhead{(kpc)} 
& \colhead{ }
& \colhead{(mag)} 
& \colhead{} 
& \colhead{(cm$^{-3}$)}   
}
\startdata
 
K59 (HP600)  &   &  16.0 & 3.6(0.2) & 0.7(0.2) & 1.5(0.2) & $<$10 \\
K68   &   &  16.1 & 4.3(0.2) & 1.3(0.2) & 1.7(0.3) & $<$10 \\
K70 (HP576,581,587,592)  &   &  16.1 & 6.3(0.3) & 2.4(0.2) & 1.7(0.1) & $<$10\\
      & a &  16.1 & 7.0(0.4) & 2.7(0.2) & 1.6(0.1) & $<$10\\
      & b &  16.1 & 6.6(0.3) & 2.5(0.2) & 2.7(0.2) & $<$10\\
      & c &  16.1 & 6.4(0.3) & 2.4(0.2) & 1.3(0.1) & 130  \\
      & d &  16.1 & 5.9(0.3) & 2.2(0.2) & 1.6(0.1) & $<$10\\
      & e &  16.1 & 7.8(0.6) & 3.0(0.2) & 1.5(0.4) & $<$10\\
K76   &   &  11.7 & 9.0(0.5) & 2.8(0.2) & 1.4(0.2) &  40  \\
K78   &   &  11.7 & 6.6(0.4) & 2.5(0.2) & 1.6(0.2) & $<$10\\
K81   &   &  11.7 & 5.2(0.3) & 1.8(0.2) & 1.8(0.2) & $<$10\\
      & a &  11.7 & 5.0(0.3) & 1.7(0.2) & 2.1(0.3) & $<$10\\
      & b &  11.7 & 5.0(0.3) & 1.7(0.2) & 1.5(0.2) & $<$10\\
K82   & a &  16.0 & 4.1(0.2) & 1.1(0.2) & 1.5(0.2) & $<$10\\
      & b &  16.0 & 3.8(0.3) & 0.9(0.2) & 1.9(0.8) & $<$10\\
K87 (HP603,609)&   &  16.0 & 4.8(0.2) & 1.6(0.2) & 1.5(0.2) & $<$10\\
      & a &  16.0 & 5.0(0.3) & 1.7(0.2) & 1.5(0.2) & $<$10\\
      & b &  16.0 & 5.7(0.3) & 2.1(0.1) & 1.5(0.1) & $<$10\\
K91 (HP613)&   &  11.6 & 5.7(0.3) & 2.1(0.2) & 1.4(0.1) &  40  \\
K92 (HP612)&   &  11.6 & 9.7(0.5) & 3.7(0.1) & 1.3(0.1) & 130  \\
K103 (HP679,680)& a &  15.8 & 3.3(0.2) & 0.4(0.2) & 1.1(0.2) & 380  \\
K132  &   &  15.7 & 4.8(0.3) & 1.6(0.2) & 1.7(0.2) & $<$10\\
K145  &   &  11.5 & 3.6(0.2) & 0.7(0.2) & 1.0(0.3) & 580  \\
K244  &   &  12.0 & 6.7(0.4) & 2.6(0.2) & 0.9(0.2) & 850  \\
K250 (HP468)&   &  12.1 & 4.5(0.2) & 1.4(0.2) & 1.4(0.1) &  40  \\
      & a &  12.1 & 4.2(0.2) & 1.2(0.1) & 1.3(0.1) & 130  \\
      & b &  12.1 & 4.8(0.3) & 1.6(0.2) & 1.4(0.2) &  40  \\
K310\da  &   &   5.4 & 3.9(0.3) & 1.0(0.2) & 1.0(0.3) & 580  \\
K314 (HP632) &   &   5.4 & 4.0(0.2) & 1.0(0.1) & 1.4(0.1) &  40  \\
K315 (HP643,649) &   &   5.4 & 3.5(0.2) & 0.6(0.1) & 1.3(0.1) & 130  \\
      & a &   5.4 & 4.3(0.2) & 1.3(0.2) & 1.0(0.1) & 580  \\
      & b &   5.4 & 3.9(0.2) & 1.0(0.2) & 1.1(0.2) & 380  \\
      & c &   5.4 & 3.7(0.2) & 0.8(0.2) & 1.3(0.1) & 130  \\
      & d &   5.4 & 3.5(0.2) & 0.6(0.1) & 1.4(0.1) &  40  \\
      & e &   5.4 & 3.1(0.2) & 0.3(0.1) & 1.3(0.1) & 130  \\
K330 (HP533) &   &   9.0 & 4.0(0.2) & 1.0(0.2) & 1.5(0.1) & $<$10\\
K343  &   &   9.0 & 4.6(0.3) & 1.5(0.2) & 1.2(0.2) & 240  \\
K353  &   &  12.2 & 7.2(0.5) & 2.8(0.2) & 1.3(0.2) & 130  \\
      & a &  12.2 & 6.0(0.3) & 2.2(0.2) & 1.4(0.1) &   40  \\
      & b &  12.2 & 7.2(0.6) & 2.8(0.2) & 2.9(1.4) & $<$10 \\
      & c &  12.2 & 5.7(0.5) & 2.1(0.2) & 0.9(0.3) &  850  \\
K391  &   &   9.0 & 4.9(0.4) & 1.6(0.3) & 0.9(0.1) &  850  \\
\tablebreak
K403 (HP549,550,556)&   &   9.0 & 3.6(0.2) & 0.7(0.1) & 1.4(0.1) &   40  \\
      & a &   9.0 & 3.9(0.2) & 1.0(0.2) & 1.1(0.1) &  380  \\
      & b &   9.0 & 3.9(0.2) & 1.0(0.2) & 1.4(0.2) &   40  \\
      & c &   9.0 & 4.3(0.2) & 1.3(0.2) & 1.3(0.2) &  130  \\
      & d &   9.0 & 3.8(0.2) & 0.9(0.1) & 1.3(0.1) &  130  \\
      & e &   9.0 & 3.3(0.2) & 0.4(0.2) & 1.3(0.1) &  130  \\
      & f &   9.0 & 3.6(0.2) & 0.7(0.2) & 1.4(0.2) &   40  \\
      & g &   9.0 & 3.4(0.2) & 0.6(0.1) & 1.4(0.1) &   40  \\
      & h &   9.0 & 3.6(0.2) & 0.7(0.1) & 1.3(0.1) &  130  \\
K414  &   &   9.1 & 3.9(0.3) & 1.0(0.2) & 1.7(0.2) & $<$10 \\
      & a &   9.1 & 5.6(1.6) & 2.1(0.9) & $<$ 4.3  & $<$10 \\
      & b &   9.1 & 3.5(0.3) & 0.6(0.2) & 1.8(0.4) & $<$10 \\
      & c &   9.1 & 3.6(0.2) & 0.7(0.2) & 1.9(0.2) & $<$10 \\
K434  &   &  10.9 & 7.3(0.4) & 2.8(0.2) & 1.5(0.4) & $<$10 \\
K442 (HP782)&   &  10.9 & 5.9(0.3) & 2.2(0.2) & 1.6(0.2) & $<$10 \\
      & a &  10.9 & 6.3(0.4) & 2.4(0.2) & 1.4(0.2) &   40  \\
      & b &  10.9 & 6.4(0.4) & 2.4(0.2) & 1.6(0.2) & $<$10 \\
      & c &  10.9 & 4.3(0.2) & 1.3(0.2) & 1.7(0.2) & $<$10\\
      & d &  10.9 & 4.3(0.3) & 1.3(0.2) & 1.6(0.2) & $<$10\\
K446\da  &   &   9.1 & 4.1(0.2) & 1.1(0.2) & 1.2(0.1) &  240 \\
K447  &   &   9.1 & 4.1(0.2) & 1.1(0.2) & 2.1(0.3) & $<$10\\
K450  & a &  12.3 & 12.3(3.1)& 4.4(0.8) & 1.3(0.4) &  130 \\
      & b &  12.3 & 3.3(0.2) & 0.5(0.2) & 0.9(0.1) &  850 \\
K461  &   &   9.3 & 3.6(0.2) & 0.7(0.2) & 1.6(0.1) & $<$10\\
K480  &   &  10.9 & 7.1(1.3) & 2.7(0.3) & 1.5(0.4) & $<$10\\
K496 (HP778)&   &  10.9 & 5.4(0.3) & 1.9(0.1) & 1.3(0.1) & 130 \\
K525 (HP625,629)&   &   9.2 & 3.7(0.2) & 0.8(0.1) & 1.5(0.1) & $<$10\\
      & a &   9.2 & 3.4(0.2) & 0.5(0.2) & 1.5(0.2) & $<$10\\
      & b &   9.2 & 4.3(0.2) & 1.3(0.2) & 1.1(0.1) & 380  \\
      & d &   9.2 & 3.2(0.2) & 0.3(0.2) & 0.8(0.4) & 1500 \\
      & e &   9.2 & 3.3(0.2) & 0.4(0.2) & 2.2(0.4) & $<$10\\
K526  &   &   9.2 & 4.7(0.3) & 1.5(0.2) & 1.3(0.1) & 130  \\
      & a &   9.2 & 4.0(0.2) & 1.0(0.2) & 1.4(0.1) & 40   \\
K531  &   &  10.8 & 6.1(0.3) & 2.3(0.2) & 1.2(0.1) & 240  \\
K536 (HP622)&   &   9.2 & 3.1(0.2) & 0.3(0.2) & 2.5(1.2) & $<$10 \\
      & a &   9.2 & 2.8(0.2) & 0.0(0.0) & 1.2(0.1) & 240  \\
      & b &   9.2 & 3.3(0.2) & 0.4(0.2) & 1.7(0.6) & $<$10\\
K536\tablenotemark{6}&   &  12.3 & 2.9(0.2) & 0.04(0.2)& 3.9(1.9) & $<$10 \\
K703 (HP721,722) &   &   9.6 & 3.0(0.2) & 0.1(0.2) & 1.3(0.1) & 130  \\
      & a &   9.6 & 4.1(0.3) & 1.1(0.2) & 1.3(0.2) & 130  \\
      & b &   9.6 & 8.6(0.7) & 3.3(0.2) & 1.6(0.3) & $<$10\\
      & c &   9.6 & 4.2(0.3) & 1.2(0.2) & 1.3(0.2) & 130  \\
      & d &   9.6 & 3.8(0.2) & 0.9(0.2) & 1.3(0.2) & 130  \\
      & e &   9.6 & 3.1(0.2) & 0.3(0.1) & 1.4(0.1) &  40  \\
      & f &   9.6 & 4.8(0.2) & 1.6(0.1) & 1.4(0.1) &  40  \\
\tablebreak
K722 (HP720,857) &   &  9.6 & 4.5(0.3) & 1.4(0.2) & 1.2(0.2) & 240  \\
      & a &  9.6 & 4.4(0.3) & 1.3(0.2) & 1.4(0.2) &  40  \\
      & b &  9.6 & 5.1(0.4) & 1.7(0.2) & 1.3(0.2) & 130  \\
K772  &   & 10.1 & 5.4(0.4) & 1.9(0.2) & 2.3(0.1) & $<$10\\
K787 (HP857) &   & 10.1 & 6.7(0.6) & 2.6(0.3) & 1.7(0.3) & $<$10\\
K838  &   & 10.3 & 7.0(0.4) & 2.7(0.2) & 1.2(0.2) & 240  \\
K851  &   & 10.4 & 3.8(0.2) & 0.9(0.1) & 1.4(0.1) &  40  \\
K856 (HP879) & a & 10.4 & 4.6(0.3) & 1.5(0.2) & 1.7(0.2) & $<$10\\
      & b & 10.4 & 4.4(0.2) & 1.3(0.2) & 1.4(0.1) & 40   \\
      & d & 10.4 & 4.5(0.2) & 1.4(0.2) & 1.4(0.1) & 40   \\
K877 (HP930,931,933)&   & 14.4 & 3.8(0.2) & 0.9(0.2) & 0.9(0.2) & 850  \\
      & a & 14.4 & 3.6(0.3) & 0.7(0.2) & 1.3(0.3) & 130  \\
      & b & 14.4 & 4.0(0.2) & 1.0(0.2) & 1.1(0.1) & 380  \\
      & c & 14.4 & 4.5(0.4) & 1.4(0.3) & 0.8(0.2) &1500  \\
      & d & 14.4 & 3.0(0.2) & 0.1(0.2) & 0.9(0.2) & 850  \\
      & e & 14.4 & 3.6(0.2) & 0.7(0.2) & 0.9(0.2) & 850  \\
K927 (HP962,964)&   & 14.2 & 3.8(0.2) & 0.9(0.2) & 1.2(0.1) & 240  \\
      & a & 14.2 & 3.8(0.2) & 0.9(0.2) & 1.3(0.1) & 130  \\
      & b & 14.2 & 3.7(0.2) & 06.8(0.2) & 1.2(0.1) & 240  \\
K931 (HP970)&   & 14.2 & 4.9(0.2) & 1.6(0.2) & 1.3(0.1) & 130  \\
K932 (HP968)&   & 14.1 & 4.1(0.2) & 1.1(0.1) & 1.3(0.1) & 130  \\
      & a & 14.1 & 4.2(0.2) & 1.2(0.1) & 1.4(0.1) &  40  \\
      & b & 14.1 & 4.0(0.2) & 1.0(0.1) & 1.3(0.1) & 130  \\
\tablenotetext{1} {Cross-references listed for objects in Pellet \etal (1978); complexes may be consistent with more than one HP object.}
\tablenotetext{2} {Radial distances taken from Walterbos \& Braun (1992).}
\tablenotetext{3} {Ratio based on observed H$\alpha$ and H$\beta$ fluxes.}
\tablenotetext{4} {Extinction determined using A(V) = 3.1E(B-V) and H$\alpha$/H$\beta$$_{int}$=2.86.}
\tablenotetext{5} {R$_{[SII]}$=6716$\rm\AA$/6731$\rm\AA$}
\tablenotetext{6} {Spectrum of K536 obtained with second position angle pointing.}
\tablenotetext{\dagger} {Spectroscopically rejected SNR candidate.}
\enddata
\end{deluxetable} 

\newpage

\begin{deluxetable}{lcrrrrrrr}
\tablecolumns{9}
\scriptsize 
\tablewidth{0pc}
\tablenum{3}
\tablecaption{SNR Nebular Conditions} 
\tablehead{
\colhead{Object\tablenotemark{1}}
& \colhead{Aperture} 
& \colhead{R$_g$\tablenotemark{2}} 
& \colhead{H$\alpha$/H$\beta$\tablenotemark{3}}
& \colhead{A$_{V}$\tablenotemark{4}} 
& \colhead{R$_{[SII]}$\tablenotemark{5}} 
& \colhead{n$_e$} \\
\colhead{ }  
& \colhead{ }  
& \colhead{(kpc)} 
& \colhead{ }
& \colhead{(mag)} 
& \colhead{} 
& \colhead{(cm$^{-3}$)}   
}
\startdata

K86   &   & 11.7 & 3.4(0.2) & 0.4(0.2)  & 1.4(0.1) &    40 \\
      & a & 11.7 & 3.4(0.2) & 0.4(0.2)  & 1.3(0.1) &   130 \\
      & b & 11.7 & 3.1(0.2) & 0.1(0.2)  & 1.4(0.1) &    40 \\
K103A & b & 15.8 & 4.0(0.2) & 0.9(0.1)  & 1.4(0.1) &    40 \\
      & c & 15.8 & 3.7(0.2) & 0.6(0.1)  & 1.3(0.1) &   130 \\
      & d & 15.8 & 4.4(0.2) & 1.1(0.2)  & 1.4(0.1) &    40 \\
      & e & 15.8 & 4.1(0.3) & 0.9(0.2)  & 1.7(0.4) & $<$10 \\
K230A &   & 12.0 & 3.4(0.2) & 0.4(0.2)  & 1.2(0.1) &   240 \\
      & a & 12.0 & 3.8(0.3) & 0.7(0.2)  & 1.2(0.1) &   240 \\
      & b & 12.0 & 4.0(0.3) & 0.9(0.2)  & 1.5(0.2) & $<$10 \\
K252 (BA22)&   &  8.7 & 3.6(0.2) & 0.5(0.1)  & 1.3(0.1) &   130 \\
      & a &  8.7 & 3.3(0.2) & 0.3(0.1)  & 1.3(0.1) &   130 \\
      & b &  8.7 & 4.0(0.2) & 0.9(0.1)  & 1.3(0.1) &   130 \\
K327  &   & 12.2 & 4.8(0.2) & 1.4(0.1)  & 1.1(0.1) &   380 \\
K425  &   & 12.2 & 3.5(0.2) & 0.4(0.2)  & 1.9(0.2) & $<$10 \\
      & a & 12.2 & 3.4(0.3) & 0.4(0.3)  & 2.1(0.4) & $<$10 \\
      & b & 12.2 & 4.0(0.3) & 0.9(0.2)  & 1.6(0.2) & $<$10 \\
K490A &   &  9.2 & 3.6(0.2) & 0.5(0.2)  & 1.6(0.1) & $<$10 \\
      & a &  9.2 & 3.5(0.2) & 0.4(0.2)  & 1.4(0.1) &    40 \\
      & b &  9.2 & 3.1(0.2) & 0.1(0.2)  & 1.6(0.2) & $<$10 \\
      & c &  9.2 & 3.6(0.2) & 0.5(0.2)  & 1.6(0.1) & $<$10 \\
K506A &   &  9.2 & 5.9(0.4) & 1.6(0.2)  & 1.4(0.1) &    40 \\
K525A (BA100)& c &  9.2 & 3.8(0.2) & 0.7(0.1)  & 1.4(0.1) &    40 \\
K526A  & b &  9.2 & 5.1(0.3) & 1.6(0.2)  & 1.4(0.2) &    40 \\
      & c &  9.2 & 8.9(0.9) & 3.3(0.3)  & 0.9(0.2) &   850 \\
K527A &   &  9.2 & 4.1(0.2) & 1.0(0.2)  & 1.7(0.2) & $<$10 \\
K594  &   & 10.7 & 7.8(0.4) & 2.9(0.2)  & 1.4(0.1) &    40 \\
K717 (BA160)&   &  9.6 & 5.9(0.4) & 1.6(0.2)  & 1.0(0.1) &   580 \\
K856A (BA212)& c & 10.4 & 4.3(0.2) & 1.1(0.1)  & 1.5(0.1) & $<$10 \\
K884 (BA650)&   & 14.4 & 4.2(0.2) & 1.0(0.2)  & 1.4(0.1) &    40 \\
      & a & 14.4 & 3.7(0.2) & 0.6(0.1)  & 1.5(0.1) & $<$10 \\
      & b & 14.4 & 5.0(0.4) & 1.5(0.2)  & 1.2(0.2) &   240 \\
K934  &   & 14.4 & 3.9(0.2) & 0.8(0.2)  & 1.5(0.1) & $<$10 \\
\tablenotetext{1} {Cross-references are listed for SNRs in Blair, Kirshner \& Chevalier (1981)
and Blair, Kirshner \& Chevalier (1982).}
\tablenotetext{2} {Radial distances taken from Walterbos \& Braun (1992).}
\tablenotetext{3} {Ratio based on observed H$\alpha$ and H$\beta$ fluxes.}
\tablenotetext{4} {Extinction determined using A(V) = 3.1E(B-V) and H$\alpha$/H$\beta$$_{int}$=3.00.}
\tablenotetext{5} {R$_{[SII]}$=6716$\rm\AA$/6731$\rm\AA$}
\enddata
\end{deluxetable} 

\newpage

\begin{deluxetable}{lcrrrrrrl}
\tablecolumns{9}
\scriptsize 
\tablewidth{0pc}
\tablenum{4}
\tablecaption{HII Region Nebular Line Ratios} 
\tablehead{
\colhead{Object\tablenotemark{1}}
& \colhead{Aperture} 
& \colhead{W$_{ap}$} 
& \colhead{E.M.}
& \multicolumn{4}{c}{Reddening Corrected Line Ratios\tablenotemark{2}}
& \colhead{ }  \\
\colhead{ } 
& \colhead{ } 
& \colhead{($^{\prime\prime}$)} 
& \colhead{H$\alpha$} 
& \colhead{[OII]/H$\beta$} 
& \colhead{[OIII]/H$\beta$} 
& \colhead{[NII]/H$\alpha$} 
& \colhead{[SII]/H$\alpha$}  
& \colhead{Morphology\tablenotemark{3}} 
}
\startdata

K59 (HP600)&  & 26.9 &  109 & 2.0(0.2) & $<$ 0.1    & 0.24(0.02) & 0.22(0.02) & diffuse      \\
K68       &   &  5.5 &  311 & 1.4(0.2) & 2.1(0.1)   & 0.54(0.03) & 0.14(0.01) & center-bright \\
K70 (HP576,581,586,587,592)&&55.9& 920 & 1.9(0.2) & 0.40(0.02) & 0.41(0.02) & 0.32(0.02) & complex \\
          & a &  7.6 & 1097 & 1.3(0.3) & 0.30(0.01) & 0.39(0.02) & 0.33(0.02) &  diffuse       \\
          & b &  9.0 & 1557 & 1.3(0.2) & 0.47(0.02) & 0.38(0.02) & 0.37(0.02) &  diffuse       \\
          & c &  9.0 & 1120 & 2.6(0.3) & 0.29(0.01) & 0.46(0.03) & 0.28(0.02) &  center-bright \\
          & d & 10.4 & 1089 & 1.6(0.2) & 0.40(0.02) & 0.35(0.02) & 0.25(0.01) &  ring           \\
          & e &  3.5 &  710 & $<$ 3.9  & $<$ 0.7    & 0.46(0.05) & 0.28(0.04) &  diffuse       \\
K76       &   &  8.3 & 1095 & $<$ 4.3  & $<$ 0.5    & 0.33(0.03) & 0.32(0.02) & center-bright  \\ 
K78       &   &  8.3 &  651 & $<$ 1.9  & $<$ 0.3    & 0.36(0.03) & 0.30(0.02) & center-bright  \\
K81       &   & 15.9 &  262 & 1.7(0.3) & $<$ 0.2    & 0.34(0.03) & 0.35(0.02) & ring           \\
          & a &  3.5 &  263 & $<$ 1.2  & $<$ 0.3    & 0.37(0.04) & 0.38(0.03) &  ring  \\
          & b &  6.9 &  227 & 2.1(0.4) & $<$ 0.3    & 0.40(0.04) & 0.38(0.03) &  ring  \\
K82       & a & 11.0 &  213 & 1.7(0.2) & $<$ 0.3    & 0.45(0.04) & 0.35(0.03) &  ring  \\
          & b &  6.9 &   85 & $<$ 1.0  & 0.44(0.03) & 0.17(0.04) & 0.17(0.04) &  ring  \\
K87 (HP603,609)&& 52.4 &561 & 2.3(0.2) & 2.5(0.1)   & 0.45(0.02) & 0.27(0.02) & center-bright+ring \\
          & a & 11.0 &  488 & 2.0(0.2) & 0.52(0.02) & 0.38(0.03) & 0.32(0.02) &  ring           \\
          & b & 15.2 & 1936 & 3.4(0.2) & 3.9(0.1)   & 0.44(0.02) & 0.25(0.01) &  center-bright \\  
K91 (HP613)&   &  9.7 &  862 & 2.5(0.2) & 0.31(0.03) & 0.46(0.02) & 0.34(0.02) & ring           \\
K92 (HP612)&   & 10.4 & 4970 & 2.7(0.3) & 0.57(0.02) & 0.47(0.02) & 0.21(0.01) & center-bright  \\ 
K103 (HP679,680)& a &  7.6 &  103 & 3.7(0.5) & 0.80(0.04) & 0.51(0.05) & 0.35(0.04) & diffuse   \\
K132      &   & 13.8 &  444 & 2.8(0.3) & 2.4(0.1)   & 0.53(0.03) & 0.27(0.02) & center-bright  \\
K145      &   &  7.5 &  152 & 0.7(0.2) & 0.22(0.01) & 0.47(0.05) & 0.22(0.03) & center-bright  \\
K244      &   &  9.0 &  519 & $<$ 2.8  & $<$ 0.5    & 0.38(0.03) & 0.31(0.03) & diffuse        \\
K250 (HP468)&& 19.3 &  723 & 4.7(0.3) & 1.39(0.05) & 0.51(0.03) & 0.38(0.02) & ring           \\
          & a & 19.3 &  347 & 7.5(0.4) & 3.0(0.1)   & 0.73(0.03) & 0.65(0.03) &  ring  \\
          & b & 29.7 &  241 & 2.7(0.3) & 0.38(0.02) & 0.39(0.02) & 0.25(0.02) &  ring  \\
K310\da   &   & 13.8 &   91 & 2.1(0.3) & 0.82(0.05) & 0.56(0.06) & 0.25(0.04) & diffuse        \\
K314 (HP632)& & 17.3 & 1250 & 1.6(0.1) & 0.13(0.01) & 0.46(0.02) & 0.29(0.01) & center-bright  \\ 
K315 (HP643,649)&& 69.7 &  341 & 2.1(0.1) & 0.37(0.01) & 0.54(0.03) & 0.22(0.01) & complex  \\
          & a &  7.6 &  284 & 1.4(0.2) & $<$ 0.3    & 0.51(0.03) & 0.34(0.03) &  diffuse       \\
          & b &  4.8 &  193 & 1.4(0.2) & 0.3(0.1)   & 0.46(0.04) & 0.33(0.03) &  diffuse       \\
          & c & 12.4 &  572 & 1.2(0.1) & 0.23(0.01) & 0.45(0.02) & 0.18(0.01) &  diffuse       \\
          & d &  9.0 &  546 & 2.7(0.1) & 0.38(0.01) & 0.64(0.03) & 0.27(0.01) &  ring           \\
          & e &  6.9 &  288 & 3.6(0.2) & 0.71(0.02) & 0.73(0.04) & 0.25(0.01) &  ring           \\
K330 (HP533)& & 19.3 &  178 & 4.6(0.3) & 1.40(0.05) & 0.76(0.05) & 0.52(0.03) & diffuse        \\
K343      &   & 14.5 &  115 & 3.9(0.5) & $<$ 0.5    & 0.5(0.1)   & 0.56(0.05) & diffuse        \\
K353      &   & 59.3 &  372 & $<$ 2.7  & 1.6(0.1)   & 0.17(0.02) & 0.30(0.02) & complex        \\
          & a & 13.1 &  561 & 2.1(0.5) & 0.95(0.04) & 0.35(0.02) & 0.31(0.02) &  ring           \\
          & b &  4.1 &  438 & $<$ 4.9  & 0.84(0.05) & 0.34(0.05) & 0.20(0.04) &  ring           \\
          & c &  6.9 &  227 & $<$ 7.0  & 2.0(0.1)   & 0.17(0.04) & 0.29(0.05) &  center-bright \\
K391      &   &  5.5 &  133 & 5.3(0.7) & $<$ 1.0    & 0.6(0.1)   & $<$ 0.4    & center-bright  \\
\tablebreak
K403 (HP549,550,556)&& 73 &  468.9 & 1.8(0.1) & 0.25(0.01) & 0.43(0.02) & 0.28(0.01) & complex  \\
          & a &  4.8 &  233 & 1.6(0.2) & $<$ 0.2    & 0.51(0.04) & 0.33(0.02) &  ring           \\
          & b &  5.5 &  271 & 1.3(0.2) & $<$ 0.2    & 0.46(0.03) & 0.37(0.02) &  ring           \\
          & c &  9.0 &  355 & 1.4(0.1) & $<$ 0.2    & 0.41(0.02) & 0.30(0.02) &  ring           \\
          & d &  9.7 &  482 & 1.6(0.1) & 0.12(0.01) & 0.47(0.02) & 0.33(0.02) &  center-bright \\
          & e &  9.0 &  300 & 1.7(0.1) & 0.26(0.01) & 0.32(0.02) & 0.28(0.01) &  diffuse       \\
          & f &  7.6 &  387 & 1.8(0.1) & 0.66(0.02) & 0.41(0.02) & 0.23(0.01) &  diffuse       \\
          & g & 11.0 & 1198 & 1.8(0.1) & 0.25(0.01) & 0.45(0.02) & 0.27(0.01) &  center-bright \\ 
          & h &  7.6 &  516 & 2.0(0.1) & 0.26(0.01) & 0.44(0.02) & 0.27(0.01) &  center-bright \\
K414      &   & 49.0 &  115 & 2.9(0.3) & $<$ 0.3    & 0.42(0.04) & 0.40(0.03) & complex        \\
          & a & 17.3 &  124 & 3.7(1.5) & $<$ 1.0    & 0.4(0.1)   & 0.2(0.1)   &  diffuse       \\
          & b & 22.8 &   53 & 2.0(0.2) & $<$ 0.2    & 0.43(0.05) & 0.39(0.05) &  diffuse       \\
          & c & 15.9 &  154 & 2.8(0.2) & 0.2(0.1)   & 0.22(0.02) & 0.40(0.03) &  ring           \\
K434      &   & 11.0 & 1886 & 3.0(0.5) & 0.91(0.04) & 0.33(0.03) & 0.12(0.01) & center-bright  \\ 
K442 (HP782)&   & 81.4 &  572 & 1.6(0.2) & 0.37(0.02) & 0.39(0.03) & 0.29(0.02) & complex        \\
          & a &  9.7 &  710 & 1.4(0.3) & 1.2(0.1)   & 0.41(0.03) & 0.27(0.02) &  ring           \\
          & b & 10.4 &  710 & 1.7(0.3) & $<$ 0.3    & 0.37(0.03) & 0.33(0.02) &  ring           \\
          & c &  9.0 &  262 & 2.0(0.2) & $<$ 0.2    & 0.42(0.03) & 0.38(0.03) &  diffuse       \\
          & d & 19.3 &  203 & 1.2(0.2) & 0.59(0.03) & 0.46(0.04) & 0.39(0.03) &  diffuse       \\
K446\da   &   &  5.5 &  577 & 4.0(0.2) & 0.48(0.02) & 0.52(0.03) & 0.24(0.02) & center-bright  \\
K447      &   & 12.4 &  176 &  4.6(0.4) & 3.3(0.1)  & 0.89(0.05) & 0.43(0.03) & diffuse        \\
K450      & a &  4.1 & 1775 & 47.4(4.5) & 3.0(0.6)  & 0.47(0.05) & 0.26(0.04) &  ring  \\
          & b & 15.9 &   77 &  1.8(0.2) & 5.1(0.2)  & 0.52(0.05) & $<$ 0.3    &  ring  \\
K461      &   &  9.0 &  464 & 1.0(0.1) & $<$ 0.05   & 0.52(0.03) & 0.29(0.02) & diffuse        \\
K480      &   &  4.8 &  456 & 3.6(1.1) & 1.01(0.04) & 0.5(0.1)   & 0.35(0.05) & ring           \\
K496 (HP778)&   & 15.9 & 1312 & 1.4(0.1) & $<$ 0.08   & 0.40(0.02) & 0.19(0.01) & center-bright  \\
K525 (HP625,629)&   & 50.4 &  535 & 2.5(0.1) & 0.61(0.03) & 0.47(0.04) & 0.38(0.02) & complex  \\
          & a &  4.2 &  189 & 1.3(0.1) & 0.34(0.02) & 0.47(0.04) & 0.28(0.02) &  ring           \\
          & b &  4.8 &  436 & 1.7(0.2) & 0.24(0.02) & 0.36(0.03) & 0.26(0.02) &  ring      \\
          & d &  4.8 &  112 & 1.0(0.2) & 0.72(0.04) & 0.47(0.05) & 0.11(0.05) &  diffuse       \\
          & e &  8.3 &  175 & 1.7(0.1) & 0.61(0.03) & 0.52(0.04) & 0.24(0.02) &  diffuse       \\
K526      &   & 53.1 &  424 & 1.0(0.2) & 0.19(0.04) & 0.43(0.03) & 0.37(0.02) & complex        \\
          & a & 24.8 &  473 & 1.2(0.1) & 0.11(0.01) & 0.38(0.02) & 0.27(0.02) &  center-bright \\
K531      &   &  2.7 & 1065 & $<$ 1.1  & 0.39(0.02) &  $<$ 0.1   & $<$ 0.1    & center-bright  \\ 
K536 (HP622)& & 11.0 &   75 & 1.4(0.2) & $<$ 0.4    & 0.46(0.04) & 0.18(0.03) & ring           \\
          & a &  4.1 &   44 & 1.3(0.2) & $<$ 0.6    & 0.24(0.05) & $<$ 0.4    &  ring  \\
          & b &  6.9 &   92 & 1.6(0.2) & $<$ 0.4    & 0.45(0.04) & 0.22(0.03) &  ring  \\
K536\tablenotemark{4}&&11.0&51&1.0(0.2)& $<$ 0.3    & 0.31(0.04) & 0.25(0.04) &  ring  \\
K703 (HP721,722)&   &102.1 &  134 & 1.8(0.1) & 1.35(0.05) & 0.32(0.02) & 0.35(0.02) & complex  \\
          & a & 10.4 &  109 & 2.2(0.3) & $<$ 0.4    & 0.6(0.1)   & 0.49(0.05) &  ring   \\
          & b & 28.3 &  520 &10.4(1.6) & 1.1(0.1)   & 0.6(0.1)   & 0.33(0.04) &  ring      \\
          & c & 12.4 &  197 & 1.8(0.3) & 0.76(0.03) & 0.48(0.04) & 0.36(0.03) &  diffuse       \\
          & d &  7.6 &  197 & 2.4(0.2) & 0.67(0.03) & 0.49(0.04) & 0.42(0.03) &  diffuse       \\
          & e &  9.0 &  492 & 2.7(0.2) & 0.91(0.03) & 0.39(0.02) & 0.19(0.01) &  center-bright \\ 
          & f & 13.8 & 2308 & 1.8(0.1) & 1.63(0.05) & 0.34(0.01) & 0.17(0.01) &  center-bright \\ 
\tablebreak
K722 (HP720,857)&   & 37 &   151.2 & 1.8(0.3) & 0.36(0.02) & 0.39(0.05) & 0.35(0.04) & diffuse  \\
          & a & 20.7 &   142 & 1.5(0.4) & 0.51(0.03) & 0.44(0.05) & 0.41(0.04) &  diffuse      \\
          & b & 16.6 &   251 & 2.0(0.4) & 0.46(0.02) & 0.4(0.1)   & 0.43(0.04) &  diffuse      \\
K772      &   & 13.1 &   206 & 6.6(1.0) & $<$ 0.7    & 0.8(0.1)   & $<$ 0.3    & diffuse       \\
K787 (HP857)&&  4.1 &   450 & $<$ 3.5  & $<$ 1.0    & 0.5(0.1)   & 0.7(0.1)   & ring          \\
K838      &   &  6.2 &  1183 & 2.2(0.6) & 0.28(0.02) & 0.51(0.04) & 0.27(0.02) & center-bright \\ 
K851      &   &  8.3 &  1035 & 3.0(0.2) & 0.45(0.02) & 0.47(0.02) & 0.24(0.01) & center-bright \\ 
K856 (HP879)& a &  4.8 &   289 & 1.5(0.2) & $<$ 0.23   & 0.46(0.04) & 0.44(0.04) &  ring    \\
          & b &  6.9 &   639 & 3.8(0.2) & 0.31(0.01) & 0.48(0.03) & 0.44(0.02) &  center-bright\\
          & d &  7.6 &   581 & 3.0(0.2) & 0.13(0.01) & 0.42(0.03) & 0.40(0.02) &  ring     \\
K877 (HP930,931,933)&& 77 &   127.9 & 4.7(0.4) & 1.7(0.1)   & 0.48(0.05) & 0.30(0.03) & complex \\
          & a &  5.5 &    93 & 2.7(0.5) & 0.52(0.03) & 0.6(0.1)   & 0.35(0.05) &  diffuse   \\
          & b & 20.0 &   269 & 5.0(0.3) & 2.6(0.1)   & 0.40(0.03) & 0.23(0.02) &  ring    \\
          & c &  4.8 &   137 & 5.8(0.7) & 2.0(0.1)   & 0.5(0.1)   & 0.38(0.05) &  ring   \\
          & d &  5.5 &    62 & 3.9(0.4) & 0.88(0.04) & 0.7(0.1)   & 0.43(0.05) &  diffuse\\
          & e &  9.7 &   134 & 5.4(0.4) & 0.73(0.03) & 0.54(0.05) & 0.30(0.03) &  diffuse \\
K927 (HP962,964)&& 26.9 &   200 & 3.2(0.2) & 0.46(0.02) & 0.42(0.03) & 0.47(0.03) & ring \\
          & a &  9.7 &   304 & 2.5(0.2) & 0.14(0.02) & 0.41(0.02) & 0.42(0.02) &  ring \\
          & b &  5.5 &   1420 & 4.0(0.3) & 0.44(0.02) & 0.44(0.04) & 0.54(0.03) &  ring \\
K931 (HP970)&& 14.5 &   592 & 2.6(0.2) & 0.24(0.01) & 0.42(0.02) & 0.24(0.01) & center-bright \\
K932 (HP968)&& 38.6 &  6973 & 1.9(0.1) & 3.2(0.1)   & 0.21(0.01) & 0.09(0.01) & complex       \\
          & a & 18.6 &  3287 & 2.6(0.1) & 1.8(0.1)   & 0.26(0.01) & 0.10(0.01) &  center-bright\\ 
          & b & 20.0 & 10283 & 1.7(0.1) & 3.6(0.1)   & 0.20(0.01) & 0.08(0.01) &  center-bright\\ 
\tablenotetext{1} {Cross-references listed for objects in Pellet \etal (1978); complexes may be consistent with more than one HP object.}
\tablenotetext{2} {Upper limits are quoted as 3$\sigma$.}
\tablenotetext{3} {Morphological classification based on Walterbos \& Braun (1992).}
\tablenotetext{4} {Spectrum of K536 obtained with second position angle pointing.}
\tablenotetext{\dagger} {Spectroscopically rejected SNR candidate.}
\enddata
\end{deluxetable} 

\newpage

\begin{deluxetable}{lcrrrrr}
\tablecolumns{7}
\scriptsize 
\tablewidth{0pc}
\tablenum{5}
\tablecaption{HII Region Line Ratios for Rare Lines} 
\tablehead{
\colhead{Object\tablenotemark{1}}
& \colhead{Aperture} & \multicolumn{5}{c}{Reddening Corrected Line
Ratios Relative to H$\beta$} \\
\colhead{ } 
& \colhead{ } 
& \colhead{[NeIII] 3868} 
& \colhead{HeI 4026}
& \colhead{HeI 4471} 
& \colhead{HeI 5876 } 
& \colhead{HeI 6678} 
}
\startdata
K068            &   & 0.7(0.1)   & \nodata    & \nodata      & \nodata       & \nodata \\
K087 (HP603,609)&   & 0.3(0.1)   & \nodata    & \nodata      & \nodata       & \nodata \\
                & b & 0.4(0.1)   & \nodata    & \nodata      & \nodata       & \nodata \\
K250 (HP468)    &   & 0.7(0.1)   & \nodata    & \nodata      & \nodata       & \nodata \\
                & a & 0.8(0.1)   & \nodata    & \nodata      & \nodata       & \nodata \\
K315 (HP643,649)& d & \nodata    & \nodata    & 0.06(0.01)   & \nodata       & \nodata \\ 
K531            &   & \nodata    & \nodata    & \nodata      & 0.38(0.02)    & \nodata \\
K932 (HP968)    &   & 0.14(0.01) & \nodata    & 0.04(0.01)   & 0.1(0.01)    & 0.03(0.01) \\
                & a & 0.1(0.1)   & \nodata    & 0.04(0.01)   & 0.1(0.01)    & 0.03(0.01) \\
                & b & 0.2(0.1)   & 0.02(0.01) & 0.04(0.01)   & 0.1(0.01)    & 0.03(0.01) \\
\tablenotetext{1} {Cross-references listed for objects in Pellet \etal (1978); complexes may be consistent with more than one HP object.}
\enddata
\end{deluxetable} 

\newpage

\begin{deluxetable}{lcrrrrrrr}
\tablecolumns{9}
\scriptsize 
\tablewidth{0pc}
\tablenum{6}
\tablecaption{Confirmed SNR Nebular Line Ratios} 
\tablehead{
\colhead{Object\tablenotemark{1}} 
& \colhead{Aperture}
& \colhead{W$_{ap}$} 
& \colhead{E.M.}
& \multicolumn{5}{c}{Reddening Corrected Line Ratios\tablenotemark{2}}\\
  \colhead{} 
& \colhead{} 
& \colhead{($^{\prime\prime}$)} 
& \colhead{H$\alpha$} 
& \colhead{[OII]/H$\beta$} 
& \colhead{[OIII]/H$\beta$} 
& \colhead{[OI]/H$\alpha$}
& \colhead{[NII]/H$\alpha$}
& \colhead{[SII]/H$\alpha$}
}
\startdata
K86   &    & 24.2 &  193 &  6.5(0.3) & 0.82(0.04) & 0.13(0.01) & 0.53(0.03) & 0.60(0.03) \\
      & a  & 15.9 &  216 &  6.1(0.3) &   1.0(0.1) & 0.13(0.01) & 0.55(0.03) & 0.60(0.03) \\
      & b  &  4.1 &  142 &  5.9(0.3) & 0.66(0.04) & 0.15(0.01) & 0.52(0.04) & 0.67(0.04) \\
K103A & b  &  7.6 &  968 &  3.6(0.2) &   1.2(0.1) &  $<$0.06   & 0.53(0.02) & 0.49(0.02) \\
      & c  &  4.1 &  769 &  8.2(0.4) &   1.5(0.1) & 0.13(0.01) & 0.77(0.03) & 0.85(0.04) \\
      & d  & 15.9 &  476 &  5.1(0.4) &   1.1(0.1) & 0.07(0.01) & 0.59(0.03) & 0.56(0.03) \\
      & e  &  4.8 &   91 &  8.4(0.9) &   0.7(0.1) &   $<$0.2   & 0.69(0.06) &   0.5(0.1) \\
K230A &    & 15.2 &  118 &  8.9(0.5) &   3.4(0.2) & 0.07(0.01) & 0.75(0.05) &   0.8(0.1) \\
      & a  &  8.3 &  186 &  9.6(0.8) &   3.4(0.2) & 0.09(0.01) & 0.73(0.04) & 0.67(0.04) \\
      & b  &  5.5 &  138 & 12.7(0.8) &   4.4(0.2) & 0.15(0.05) & 0.78(0.06) &   0.8(0.1) \\
K252 (BA22)&    &  8.9 &  683 &  9.5(0.4) &   2.4(0.1) & 0.21(0.01) & 1.07(0.05) &   1.1(0.1) \\
      & a  &  4.8 &  467 &  9.5(0.4) &   2.7(0.1) & 0.25(0.01) & 1.12(0.05) & 1.13(0.05) \\
      & b  &  4.1 & 1006 & 10.7(0.5) &   2.4(0.1) & 0.18(0.01) & 1.04(0.05) & 1.05(0.04) \\
K327  &    &  7.6 & 1065 &  6.7(0.4) &   1.6(0.1) & 0.25(0.01) &   1.2(0.1) & 1.02(0.04) \\
K425  &    &  9.7 &   58 &  9.3(0.7) & 0.83(0.05) & 0.22(0.04) & 0.81(0.07) &   0.9(0.1) \\
      & a  &  4.1 &   37 &  9.9(1.0) &   $<$0.74  &   0.5(0.1) &   0.7(0.1) &   1.0(0.1) \\
      & b  &  4.1 &   95 & 10.5(0.8) &   1.0(0.1) &   $<$0.2   & 0.80(0.07) &   1.0(0.1) \\
K490A &    & 42.1 &  308 &  5.2(0.3) & 0.76(0.04) & 0.10(0.05) & 0.75(0.04) & 0.68(0.04) \\
      & a  & 15.2 &  291 &  6.8(0.4) & 0.97(0.05) &   $<$0.25  & 0.90(0.05) & 1.05(0.05) \\
      & b  &  6.9 &  241 &  4.5(0.3) & 0.87(0.04) &   $<$0.19  & 0.63(0.03) & 0.45(0.03) \\
      & c  &  9.7 &  431 &  3.6(0.2) & 0.31(0.02) &   $<$0.06  & 0.62(0.03) & 0.54(0.03) \\
K506A &    &  7.6 &  452 &  3.5(0.4) &   1.5(0.1) &   $<$0.07  & 0.58(0.03) & 0.62(0.05) \\
K525A (BA100)& c  & 20.0 & 1090 &  2.9(0.1) & 0.66(0.03) & 0.03(0.01) &  0.5(0.02) & 0.45(0.02) \\
K526A  & b  &  8.3 &  325 &  2.7(0.3) & 0.52(0.10) &   $<$0.16  & 0.55(0.04) & 0.76(0.05) \\
      & c  &  3.5 &  646 &    $<$2.7 &   $<$1.1   &   $<$0.43  & 0.29(0.05) &   0.6(0.1) \\
K527A &    & 23.5 &  33 &  2.3(0.2) & 0.29(0.02) & 0.23(0.02) & 0.48(0.03) & 0.42(0.03) \\
K594  &    &  8.3 & 2189 &  8.4(0.6) &   0.9(0.1) & 0.29(0.01) & 1.07(0.05) & 1.16(0.05) \\
K717 (BA160)&    & 11.7 & 1107 &  6.2(0.4) & 4.0(0.2)   & 0.19(0.01) & 0.95(0.05) & 0.88(0.04) \\
K856A (BA212)& c  &  6.9 &  675 &  4.8(0.3) & 0.12(0.01) & 0.08(0.01) & 0.54(0.03) & 0.68(0.03) \\
K884 (BA650)&    & 34.5 &  220 &  6.3(0.4) & 0.09(0.03) & 0.10(0.01) & 0.48(0.03) & 0.53(0.03) \\
      & a  &  6.2 &  434 &  4.7(0.3) & 0.15(0.01) &   $<$0.06  & 0.50(0.03) & 0.45(0.02) \\
      & b  &  3.5 &  213 &  9.3(0.9) & 0.6(0.1)   &   $<$0.36  & 0.50(0.06) &   0.7(0.1) \\
K934  &    & 17.9 &  191 &  5.0(0.3) & 0.41(0.02) & 0.10(0.01) & 0.52(0.03) & 0.60(0.03) \\
\enddata
\tablenotetext{1}  {Cross-references are listed for SNRs in Blair, Kirshner \& Chevalier (1981) 
and Blair, Kirshner \& Chevalier (1982).}
\tablenotetext{2} {Upper limits are quoted as 3$\sigma$.}
\end{deluxetable} 

\newpage

\begin{deluxetable}{lcrr}
\tablecolumns{4}
\scriptsize 
\tablewidth{0pc}
\tablenum{7}
\tablecaption{SNR [OIII] Temperatures} 
\tablehead{
\colhead{Object\tablenotemark{1}} 
& \colhead{Aperture}
& \colhead{$\frac{I(4959+5007)}{I(4363)}$} 
& \colhead{T$_e$ (K)\tablenotemark{2}} 
}
\startdata
 K086       & a &  12.4(3.0)  &   58,500(15,000) \\
 K230A      &   &  16.9(1.0)  &   39,000(3,400) \\
            & a &  12.9(0.7)  &   55,700(4,000) \\
 K252 (BA22)&   &  18.0(0.9)  &   37,000(3,000) \\
            & a &  17.5(0.9)  &   38,000(2,000) \\
            & b &  18.0(0.9)  &   37,000(3,000) \\
\tablenotetext{1} {Cross-references are listed for SNRs in Blair, Kirshner \& Chevalier (1981) and Blair, Kirshner \& Chevalier (1982).}
\tablenotetext{2} {Based on Kaler \etal (1976) formalism.}
\enddata
\end{deluxetable} 

\newpage

\begin{figure}
\figurenum{1a}
\epsscale{0.8}
\caption{Slit positions and object aperture centers for all discrete 
objects and substructures in the 2D spectra.  Grey-scale saturation
levels in emission measure (pc cm$^{-6}$) are given at the lower right
hand corner of each image; image scaling is linear to bring out
variations most dramatically.  Images are 78'' by 78''.  Complete
images have been published in Walterbos \& Braun (1992).  }
\end{figure}

\newpage
\begin{figure}
\figurenum{1b}
\epsscale{0.8}
\caption{Slit positions and object aperture centers for all discrete 
objects and substructures in the 2D spectra.  Grey-scale saturation
levels in emission measure (pc cm$^{-6}$) are given at the lower right
hand corner of each image; image scaling is linear to bring out
variations most dramatically.  Images are 78'' by 78''.  Complete
images have been published in Walterbos \& Braun (1992).  }
\end{figure}

\newpage
\begin{figure}
\figurenum{2}
\epsscale{0.8}
\plotone{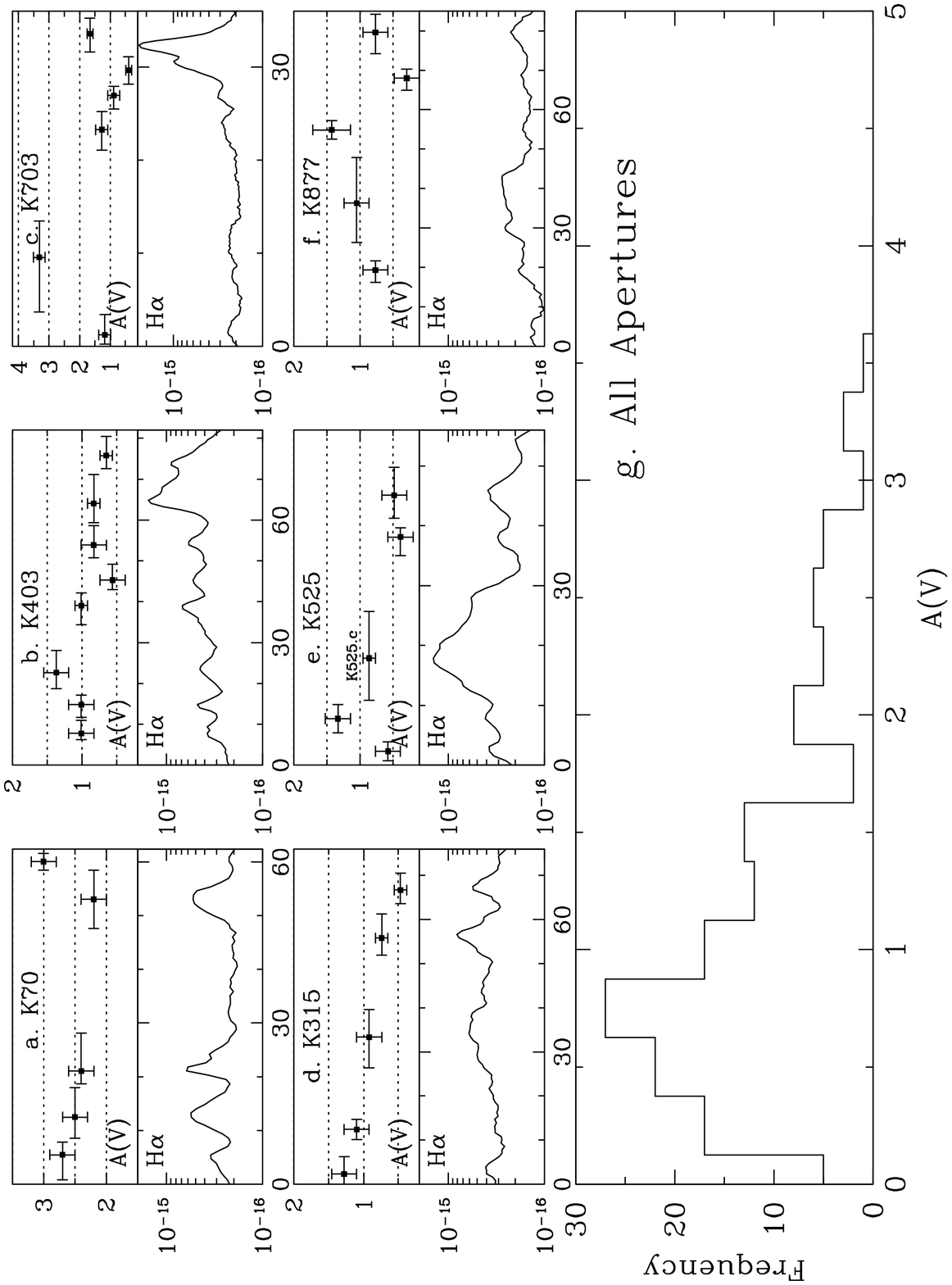}
\caption{Extinction properties derived from the Balmer 
decrements of our complete sample of M31 nebulae.  The separate panels
in (a-f) show the variation of A(V) (in magnitudes) across the spatial
dimension (in arcseconds) of large HII region complexes along with the
H$\alpha$ profile along the spectrum slit.  Each point corresponds to
a separate aperture and the bar delineates the section of the spectrum
included in each aperture.  An embedded SNR candidate in the HII
region complex, K525, is identified.  Dashed lines delineate the
horizontal scale and are drawn only as a reference.  Panel (g) shows
the distribution of extinctions found in all the HII region and SNR
spectra.}
\end{figure}

\newpage
\begin{figure}
\figurenum{3}
\epsscale{0.8}
\plotone{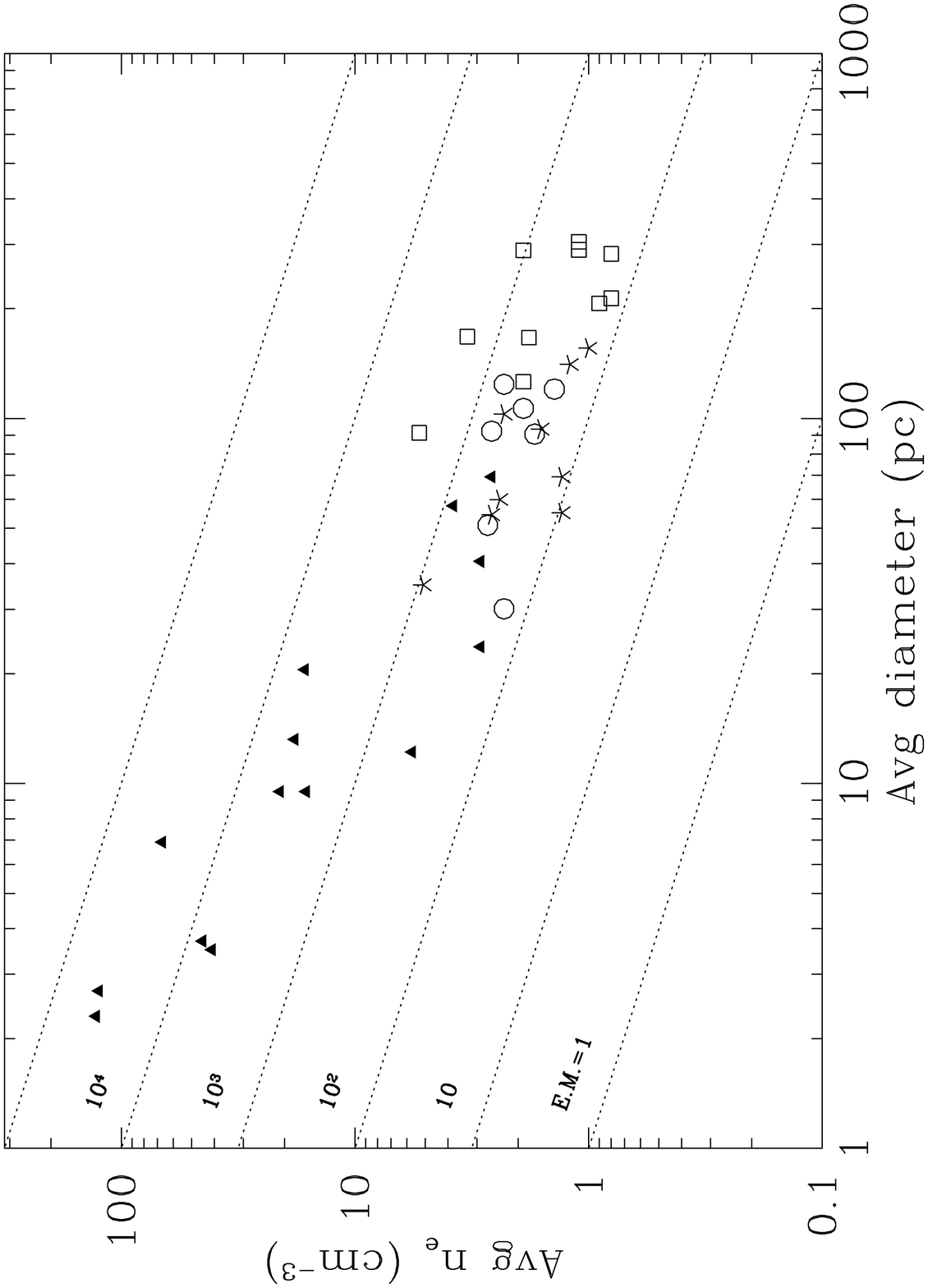}
\caption{A comparison of the size distribution and
electron densities of the HII regions in our sample.  This figure
shows average rms-electron density (estimated from the H$\alpha$
emission measures assuming unit filling factor) plotted against
average nebular diameters.  Dotted lines represent lines of constant
emission measure in pc cm$^{-6}$.  Point types specify morphological
classification adopted from Walterbos \& Braun (1992):
center-brightened sources (triangles), ring-like nebulae (open
circles), diffuse nebulae (stars) and complexes (open squares).}
\end{figure}

\newpage
\begin{figure}
\figurenum{4}
\epsscale{0.8}
\plotone{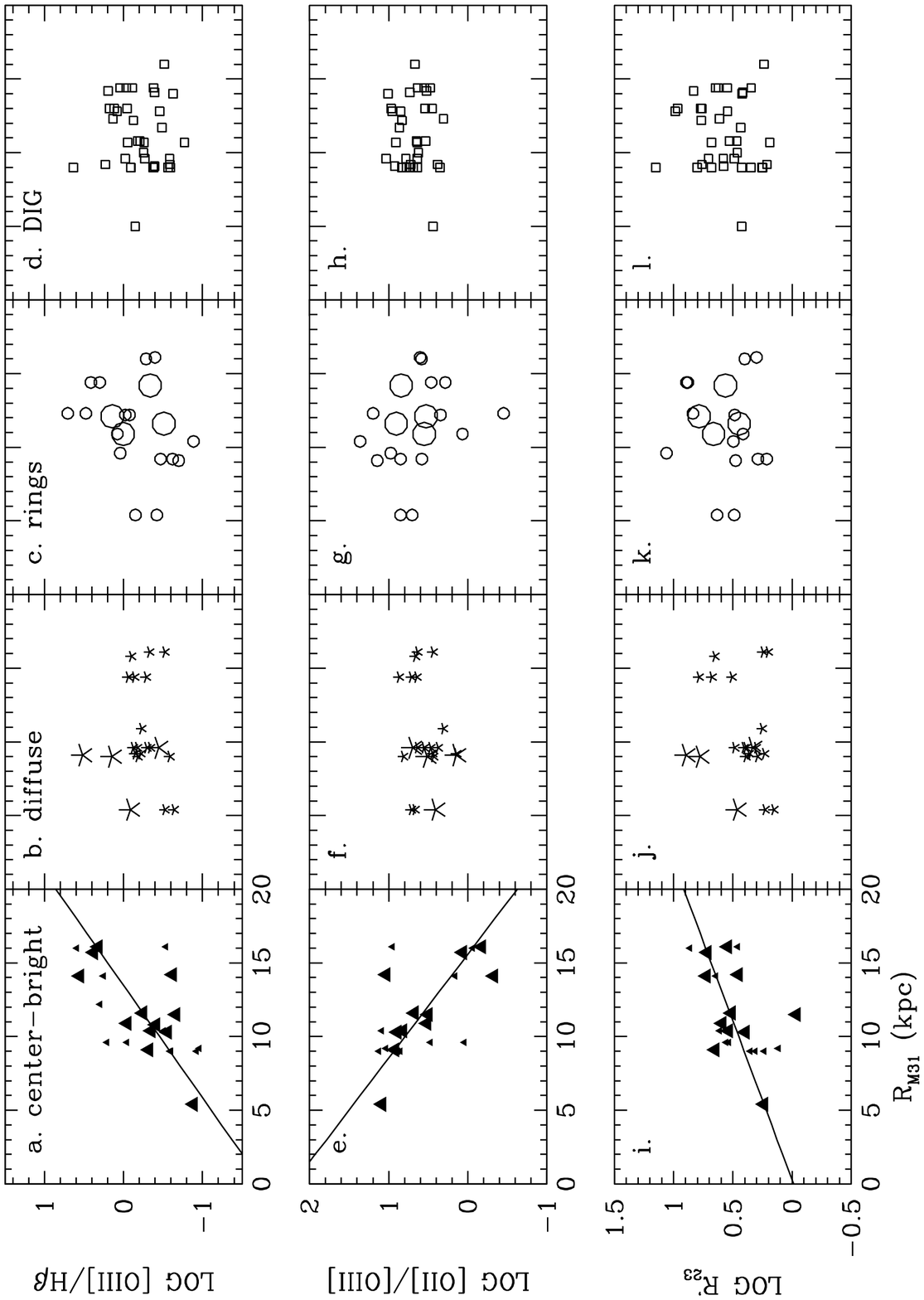}
\caption{Radial distributions of the detected oxygen line 
ratios, [OIII]/H$\beta$ ({\bf top row}), [OII]/[OIII] ({\bf middle
row}), and R$_{23}$ ({\bf bottom row}) for HII regions and DIG.
Panels demonstrate ({\it from left to right}) line variations with
galaxy position for individual center-brightened nebulae and knots
that are part of larger complexes (large and small triangles,
respectively); for diffuse nebulae and diffuse sections of complexes
(large and small stars, respectively); and for rings and arcs in
complexes (large and small open circles, respectively).  The regions
of DIG are represented by open squares.  Errorbars (not plotted) are
approximately the size of the large points.  Least squares linear fits
are plotted here only for those cases where the correlation
coefficient $\ge$ 0.5.  Radial gradients which are expected in these
line ratios due to the chemical abundance gradient in the galaxy, only
appear in the center-bright HII regions (left-most panels).}
\end{figure}

\newpage
\begin{figure}
\figurenum{5}
\epsscale{0.8}
\plotone{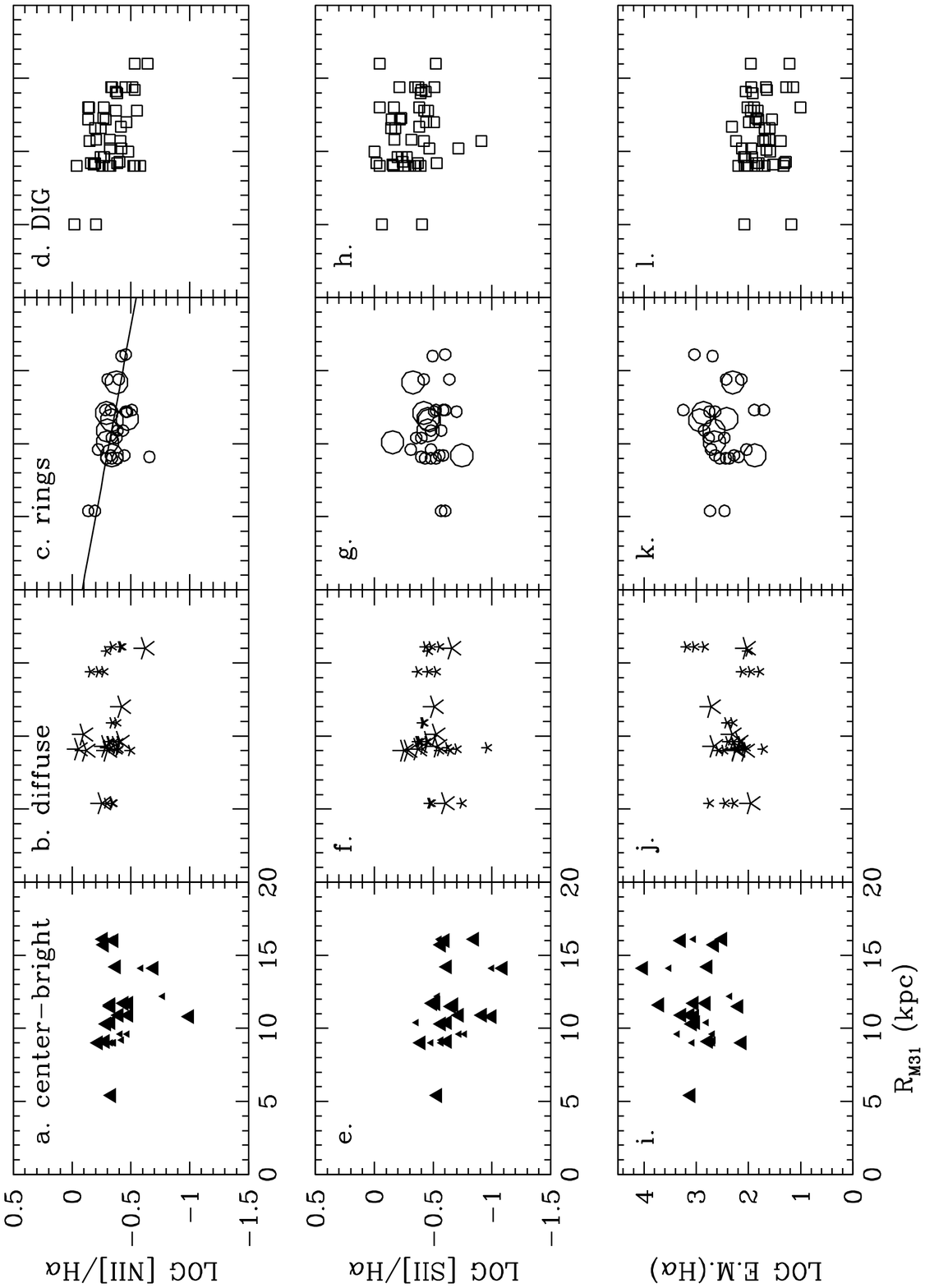}
\caption{Radial distributions for the detected line ratios
[NII]/H$\alpha$ ({\bf top row}), [SII]/H$\alpha$ ({\bf middle row}),
and the H$\alpha$ emission measure obtained in each spectrum ({\bf
bottom row}) for HII regions and DIG.  Point types and panels are the
same as in Fig 4.  The only significant linear correlation was found
in the [NII] emission of ring nebulae.}
\end{figure}

\newpage
\begin{figure}
\figurenum{6}
\epsscale{0.8}
\plotone{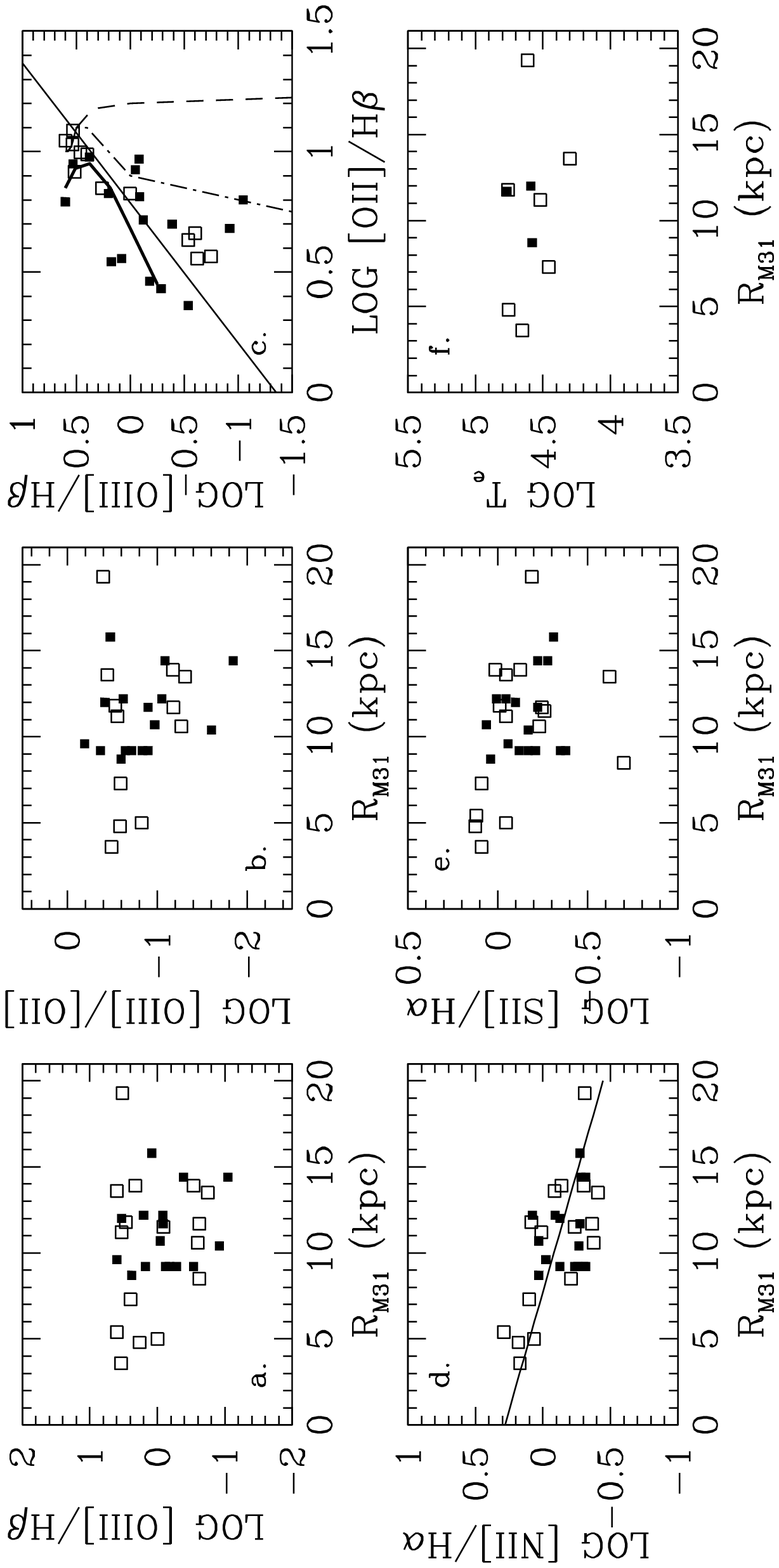}
\caption{Radial distribution of line ratios and physical
conditions for the confirmed SNRs from this paper (filled squares) and
SNRs from Blair, Kirshner \& Chevlier (1981, 1982) (open squares).
Linear $\chi$$^{2}$ fits are shown where applicable.  {\bf Top row}:
radial variation in the line ratios (a) [OIII]/H$\beta$ and (b)
[OIII]/[OII], and the correlation between the [OIII] and [OII]
emission in (c).  Shock-ionization models from Dopita \etal (1984) are
plotted as follows: dash-dot and dash curves show predictions for two
models (standard and pre-ionized) for a range of shock velocity from
50 to 140 km sec$^{-1}$; thick solid curve shows the predicted effects
on the [OII] and [OIII] emission for a variation in oxygen metallicity
from N(O) = 2 x 10$^{-3}$ (highest point) to 8 x 10$^{-4}$.  
{\bf Bottom row}: radial variation in the line ratios (d) [SII]/H$\alpha$, 
(e) [NII]/H$\alpha$, and the electron temperature determined from the 
[OIII] line ratio.  Fits are plotted only for significant correlations 
as in previous plots.}
\end{figure}

\newpage
\begin{figure}
\figurenum{7}
\epsscale{0.8}
\plotone{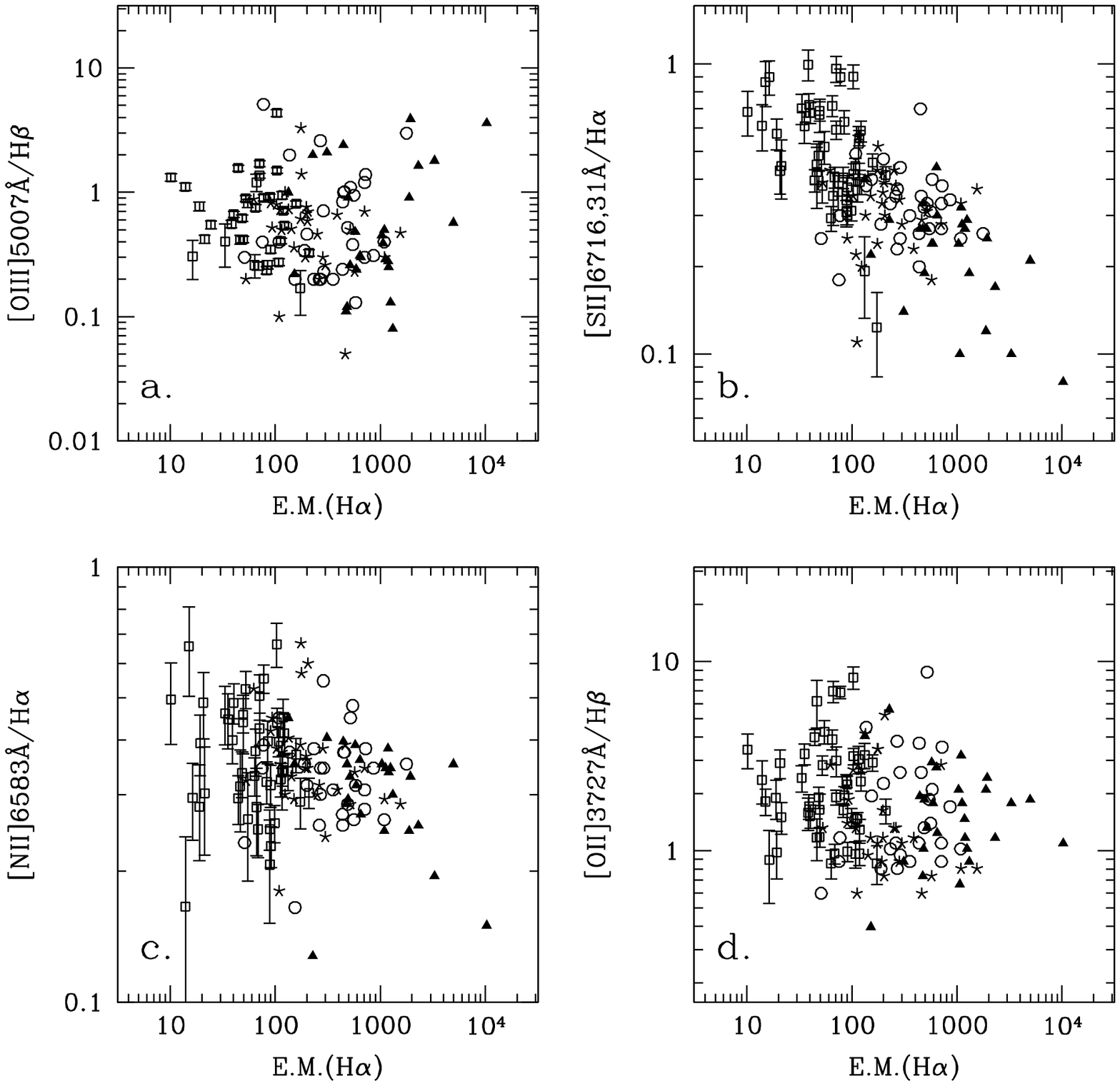}
\caption{Line ratio variations with H$\alpha$ emission
measure for HII regions and DIG.  Point types are the same as in Fig
5.  The errorbars have only been plotted for the DIG to minimize
clutter; the errorbars on the HII region data are smaller.}
\end{figure}

\newpage
\begin{figure}
\figurenum{8}
\epsscale{0.6}
\plotone{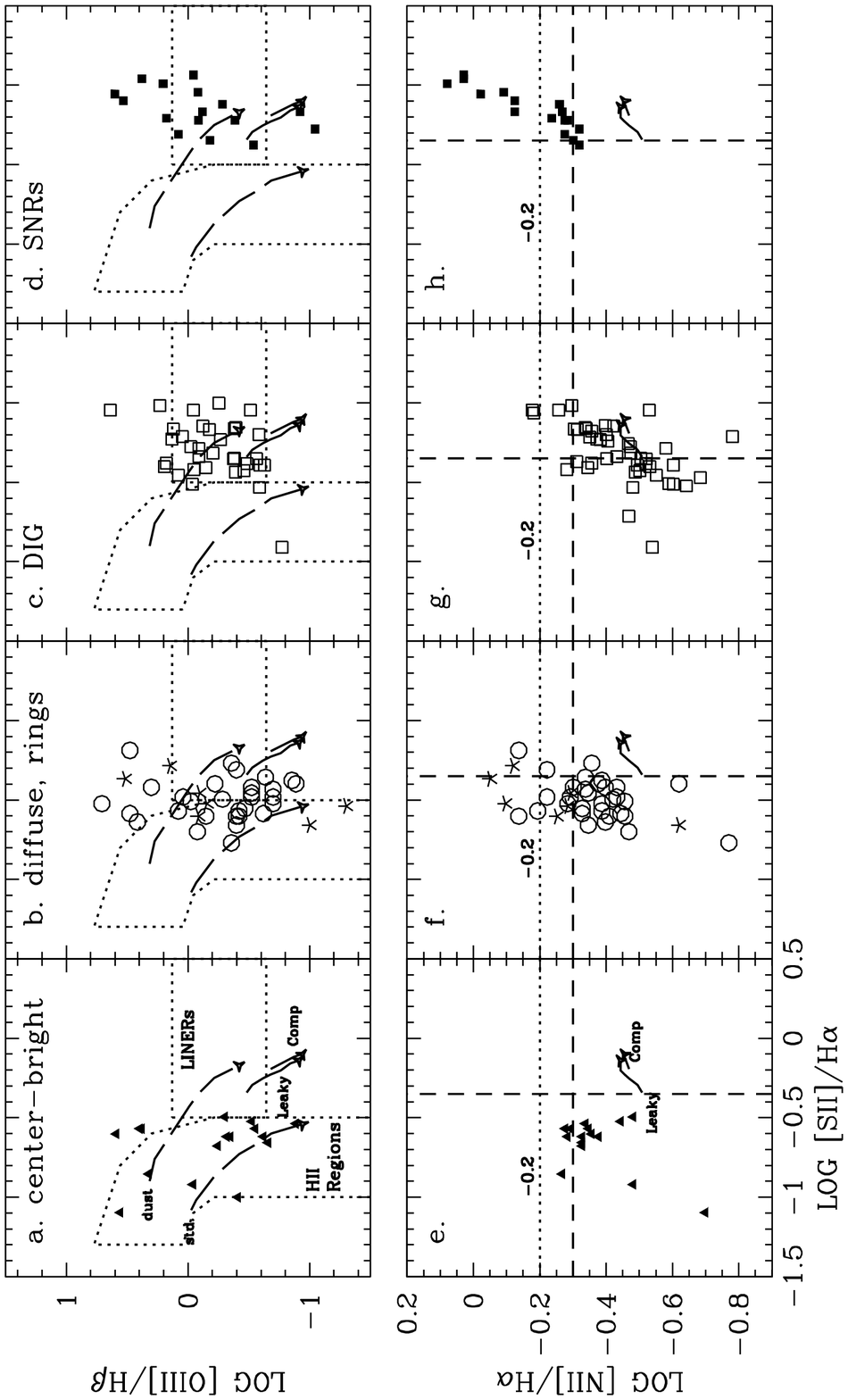}
\caption{Diagnostic diagrams based on the [OIII], [NII],
and [SII] emission in the various nebulae and DIG.  {\bf Top row}:
Theoretical models from Shields \& Filippenko (1990) are labelled as
outlined boxes and separate photoionized objects (labelled ``HII
regions'') from shock-ionized objects (labelled ``LINERs'').  A smooth
transition in [SII] emission links the brightest HII regions (leftmost
panel), fainter HII regions, DIG and SNR (rightmost panel).  Two
photoionization models of Sokolowski (1993) shown as dashed lines are
labeled ``std'' (standard model) and ``dust'' (dust-depletion model)
in the first panel.  The predictions of both models reproduce the
increase in [SII] and decrease in [OIII] seen between the brightest
HII regions and the DIG.  The leaky HII region and composite
photoionization models of Domg\"orgen \& Mathis (1994) are also
plotted as solid curves and labelled ``Leaky'' and ``Comp''; the
models are plotted from log(q)=-4 to log(q)=-2, where q is the
ionization parameter as defined by Domg\"orgen \& Mathis (1994).  The
arrows on the curves indicate the direction of decreasing ionization
parameter.  {\bf Bottom row}: A comparison of the [SII] and [NII]
emission properly distinguishes photoionized and shock-ionized
objects.  The two dashed lines deliniate the limiting values which
best separate the two ionization mechanisms: [SII]/H$\alpha$=0.45 and
[NII]/H$\alpha$=0.5.  Clearly, the confirmed SNRs lie in the upper
right quadrant while the HII regions and DIG mostly lie outside of
this region.  Solid lines represent the two photoionization models of
Domg\"orgen \& Mathis (1994) as indicated above.  The dotted line at
log([NII]/H$\alpha$)=-0.2 is a lower limit based on shock ionization
models and is used as a reference for the discussion of Fig 9.}
\end{figure}

\newpage
\begin{figure}
\figurenum{9}
\epsscale{0.6}
\plotone{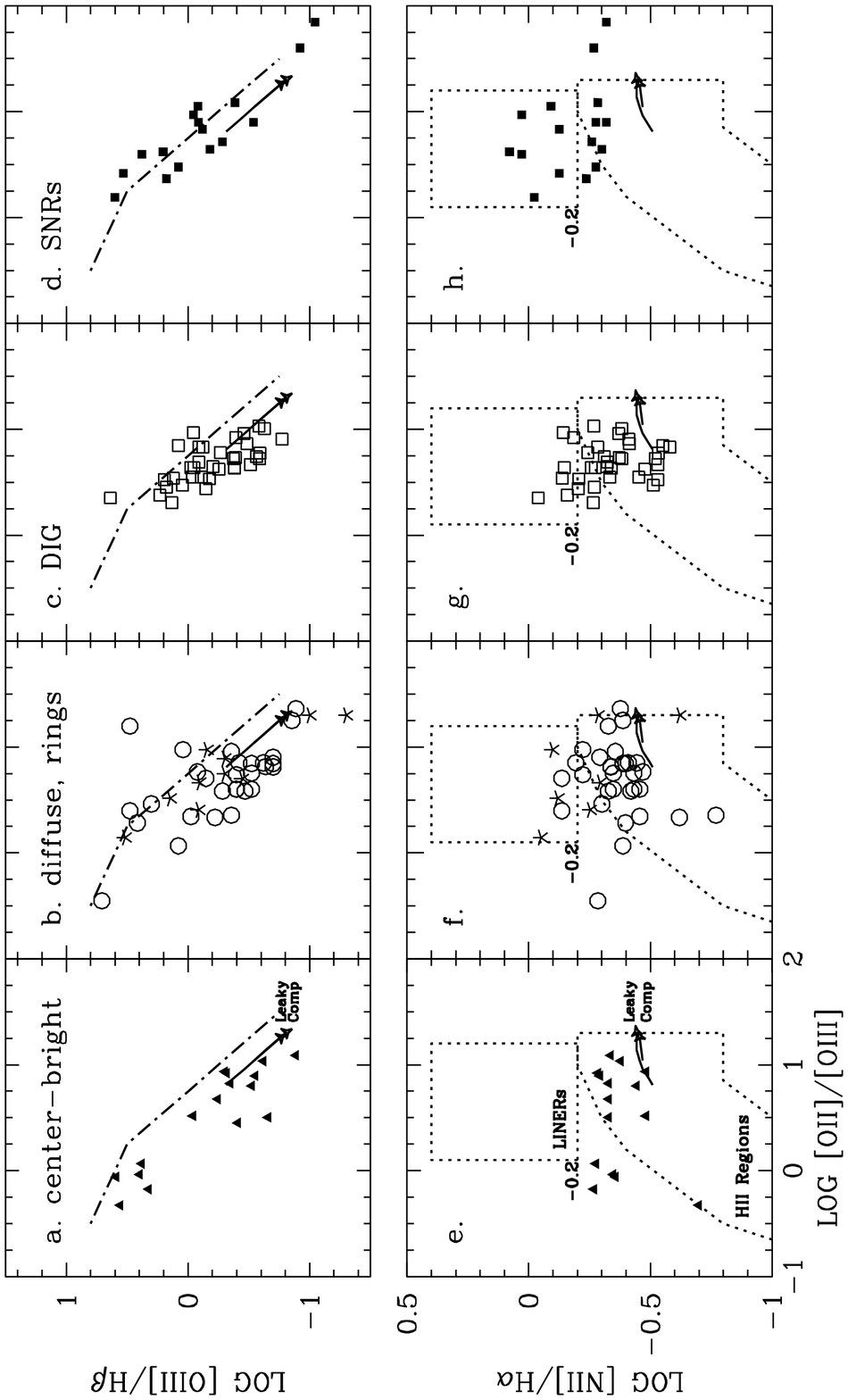}
\caption{Diagnostic diagrams for various types of HII 
regions, DIG and SNRs in terms of [OIII], [NII], and the ionization
line ratio [OII]/[OIII].  {\bf Top row}: All HII region points lie
below the upper envelope of theoretical photoionization models (Evans
\& Dopita, 1985) marked as the dash-dot curve.  The SNR data lie mostly
close to or above this curve while the DIG points lie beneath,
indicating that the [OIII] emission from the DIG is consistent with
photoionization.  The photoionization models of Domg\"orgen \& Mathis
(1994) are also plotted and labelled as in Fig 8.  {\bf Bottom
row}: [NII] emission relative to [OII]/[OIII] is also used to separate
photoionization from shock-ionization; the outlined boxes represent
Shields \& Filippenko (1990) theoretical models for photoionized and
shock-ionized gas.  The HII region points fit mostly within the
photoionization regime while the SNR close to or inside the shock
ionization box.  Many of the SNRs fall below the lower limits on the
[NII] emission for shock ionization (log([NII]/H$\alpha$)=-0.2);
however, this is likely an abundance effect.  These same data points
have high [SII] consistent with shock ionization as shown in Fig 8h.
Again the DIG points follow close along with the HII reigons,
demonstrating the [NII] emission from DIG is mostly consistent with
photoionization.}
\end{figure}

\end{document}